\documentclass[aps,prf,preprint,groupedaddress,longbibliography]{revtex4-1}

\usepackage{xcolor,bm}
\usepackage{bm,mathtools,esint}
\usepackage{graphicx,booktabs}
\usepackage{subfig,floatrow}
\usepackage{cleveref}

\begin{document}

%\preprint{}

\title{Macroscopic forcing method: a tool for turbulence modeling and analysis of closures}
%\title{Scaled up operators for continuum transport equations using a macroscopic forcing method}


\author{Ali Mani}
\email{alimani@stanford.edu}
\author{Danah Park}
%\author{Ali Mani}
\affiliation{Department of Mechanical Engineering, Stanford University,
Stanford, CA 94305, USA}
%\thanks{}


\begin{abstract}
This study presents a numerical procedure, which we call the macroscopic forcing method (MFM), which reveals the differential operators acting upon the mean fields of quantities transported by underlying fluctuating flows. Specifically, MFM can precisely determine the eddy diffusivity operator, or more broadly said, it can reveal differential operators associated with turbulence closure for scalar and momentum transport. We present this methodology by considering canonical problems with increasing complexity. Starting from the well-known problem of dispersion of passive scalars by parallel flows we elucidate the basic steps in quantitative determination of the eddy viscosity operators using MFM. Utilizing the operator representation in Fourier space, we obtain a stand-alone compact analytical operator that can accurately capture the non-local mixing effects. Furthermore, a cost-effective generalization of MFM for analysis of non-homogeneous and wall-bounded flows is developed and is comprehensively discussed through a demonstrative example. Extension of MFM for analysis of momentum transport is theoretically constructed through the introduction of a generalized momentum transport equation. We show that closure operators obtained through MFM analysis of this equation provide the exact RANS solutions obtained through averaging of the Navier-Stokes equation. We introduce MFM as an effective tool for quantitative understanding of non-Boussinesq effects and assessment of model forms in turbulence closures, particularly, the effects associated with anisotropy and non-locality of macroscopic mixing.   
\end{abstract}

%\keywords{}

\maketitle

\section{Introduction}\label{sec:introduction}
The kinetic theory of gases has been an invaluable tool for developing continuum models governing transport of mass, momentum, and energy. Without such models, prediction of flows and heat transfer in natural and industrial systems would have required tracking of individual fluid molecules by solving an enormous system of coupled molecular dynamics equations. To date, the largest of such calculations can only span domains of size of order $100nm$ which is by far much smaller than many engineering scales. Kinetic theory bridged this gap by analyzing interactions of molecules in a statistical sense. These analyses provide transport coefficients to partial differential equations (PDEs) describing transport phenomena in a macroscopic sense\cite{Vincenti1965}. Unlike the underlying chaotic Brownian dynamics, these scaled-up Eulerian PDEs result in smooth solutions with characteristic length scales multiple orders of magnitude larger than molecular scales. These PDEs therefore offer drastic computational savings by lowering the required number of degrees of freedom in the systems to be solved. 

The diffusion equation and the Navier-Stokes equation governing the evolution of mass, heat, and momentum are examples of such transport PDEs. These equations however, do not completely close the gap of length scales all the way to the engineering scales. Most notably, as many engineering applications involve flows at high Reynolds numbers, the Navier-Stokes equation itself admits chaotic solutions spanning a wide spectrum of length scales\cite{Moin1997}, rendering the system-level computations prohibitively expensive. Again, as primary quantities of interest involve only the averaged fields, researchers have actively sought approaches for further scale up of the mathematical models over the past decades.  

Central in this context is the RANS (Reynolds-Averaged Navier-Stokes) methodology \cite{Pope2001,Speziale1991,Speziale1998,Durbin2018}. The goal of RANS models is to predict the average velocity field in a turbulent flow without the need of tracking individual chaotic eddies. In this case, the average is the mean over many ensembles of realizations of the same system under the same boundary conditions. However, as most systems involve statistically time-stationary turbulence, temporal averages are equivalent to ensemble averages\cite{Pope2001}. Both averages result in highly smooth mean fields with a much smaller number of degrees of freedom suitable for pursuing scaled-up models. 

RANS modeling has, from its inception, been inspired by approaches from kinetic theory. Most notably, the majority of the RANS models\cite{Spalart1992,Hanjalic1972,Wilcox2008,Menter1994,Durbin1995} are based on the Boussinesq approximation\cite{Boussinesq1877} suggesting closures in terms of diffusion operators with the coefficient referred to as eddy diffusivity.  By making an analogy to the kinetic theory, this approximation implies that in the same way that mixing by Brownian motion leads to a scaled-up continuum model involving a diffusion equation with an isotropic coefficient tensor, turbulent mixing should be expressible in terms of diffusion operators when statistically averaged. However, unlike such Brownian motion, turbulent fluctuations are often non-isotropic in practice.  Aside from this issue, a major criticism of this analogy is that the diffusion approximation requires separation of scales between the scaled-up fields and the underlying microscopic, sub-scale, physics. This is the case for molecular-to-continuum modeling since the continuum scale is often orders of magnitude larger than the mean free path of molecules. However, in turbulent flows the underlying chaotic eddies span a wide range of scales; large eddies, which have the highest statistical significance, are often as large as the scale of the mean fields. Therefore, the Boussinesq approximation, while being useful for qualitative analysis and scaling assessments, is not accurate for quantitative predictions.

Despite these shortcomings, the majority of turbulence models in use employ eddy-diffusivity closures based on the Boussinesq approximation\cite{Spalart1992,Hanjalic1972,Wilcox2008,Menter1994,Durbin1995}. Among the models that do not use the Boussinesq approximation, some still retain diffusion-type operators in which the mean momentum flux is expressed as a function of local velocity gradients\cite{Pope1975}. Even the models that do not use any closure at the momentum transport level, such as the Reynolds stress closure models\cite{Launder1975}, still retain similar approximations in closure of higher moments. Model-form inaccuracies in turbulence models are to some degree compensated by tuning of the model coefficients and tailoring them for specific regimes and applications. As a result, turbulence models are still far from being universal and truly predictive. 

%Another related issue is the problem of verification of RANS models. Verification practice today is done by comparing the solution of the RANS models against a reference solution. For example, one verification test case is to determine whether a RANS model quantitatively captures the log-layer for wall bounded turbulent flows (** cite a reference**). Fact finding via such comparisons can be corrupted by the tuning procedure. Since eddy-diffusivity, as the model coefficient, can be arbitrarily defined in space, it is often possible to tune an inaccurate model form to produce the desired solution field. True verification should not only be able to certify the solution field produced, but also the model form and its coefficients. 

In this report we present a method, which we call the macroscopic forcing method (MFM) that allows determination of the ``closure operators" that govern the mean-field-mixing by any underlying flow field. MFM acts similar to the way that molecular dynamics simulations reveal transport coefficients for a continuum model, with the exception that in MFM both input and output spaces are continua and that MFM does not make any simplifying assumption such as isotropy or separation of scales between turbulence and mean fields (scaled up fields). MFM precisely reveals the degree of non-locality and anisotropy of the differential operators governing the mean transport. %Understanding these ingredients provides an unprecedented advancement in the development of universal turbulence models.

%In the midst of the challenge of determining accurate RANS models, decades ago, supercomputers allowed direct numerical simulation (DNS) of canonical turbulent flows\cite{Kim1987} and provided unprecedented access to their detailed underlying transport processes. Despite drastic improvements in computational power, DNS continues to remain inaccessible to industrial applications, and is only viable for canonical and often academic settings. Since its advent, it has been expected that understanding based on DNS will inspire novel RANS models with superior accuracy compared to those developed prior to the DNS era. However, thus far DNS, at best, has been used to provide reference mean fields for the tuning of RANS model coefficients, while the model form has been set a priori. A missed opportunity has been the use of DNS for revealing the model form itself. MFM closes this gap by combining DNS with a novel statistical analysis. 

Lastly, the approach, which we shall introduce, unifies the problem of macroscopic modeling between low-Reynolds-number laminar flows and high-Reynolds-number turbulent flows. As a starting point, we consider the problem of dispersion of passive scalars by a laminar parallel flow, first introduced by G. I. Taylor\cite{Taylor1953}. By recasting this problem in wavenumber space, we show that while MFM correctly reproduces Taylor's approximation for the low wavenumber limit, it provides a natural framework for extending his solution to all wavenumbers. The paper then continues to pave the way for the development of MFM for more complex flows, involving statistical inhomogeneity and wall boundary conditions. Lastly, an extension of MFM to the analysis of momentum transport is theoretically formalized by introducing a microscopic equation representing momentum transport in the generalized sense, and providing a theoretical proof regarding convergence of the statistical solution of this equation to the solution obtained from Reynolds-averaging of the Navier-Stokes equation. The paper ends by providing useful insights about MFM from both physical and mathematical point of views.

%We then report that the application of MFM to homogeneous turbulence, results in similar operators as those obtained from analysis of parallel flows.  

\subsection{Problem Statement} 
The purpose of this section is to introduce the basic terminology and mathematical notation used throughout the paper. While we ultimately develop a general approach that can treat a wide range of transport phenomena, in this section we introduce the problem by considering the passive scalar advection-diffusion equation. Later, we introduce extensions to momentum fields governed by the Navier-Stokes equation. The starting point is the continuum microscopic equation: 
\begin{equation}
\mathcal{L}c\left(x_1,...,x_n,t\right)=0,
\label{eq:microscopic}
\end{equation}
where $c$ represents the transported field, while $x_1$ to $x_n$ and $t$ are the independent variables, which correspond to the spatial and temporal coordinates, respectively. $\mathcal{L}$ is a differential operator representing the transport physics. For example, when the advective flow is incompressible, one can write
\begin{equation}
\mathcal{L}=\frac{\partial}{\partial t}+u_j\frac{\partial}{\partial x_j}-\frac{\partial}{\partial x_j}\left(D_{M}\frac{\partial}{\partial x_j}\right), 
\label{eq:advdiff}
\end{equation}
where $u_j\left(x_1,...,x_n,t\right)$ represents the advecting velocity field, and $D_{M}$ represents molecular diffusion. Equation~(\ref{eq:microscopic}) is accompanied by initial and boundary conditions. When~(\ref{eq:microscopic}) is written in discretized form, these conditions can be embedded in the operator $\mathcal{L}$ and the right hand side of (\ref{eq:microscopic}). 

We refer to Equation~(\ref{eq:microscopic}) as the ``microscopic equation'' or the ``microscopic transport equation," where $\mathcal{L}$ is called the microscopic operator acting on the microscopic field $c$. Next, we define the ``macroscopic" or scaled-up field, $\overline{c}$, which represents the quantity of engineering interest. Mathematically, $\overline{c}$ is defined as the average of $c$ over a certain subset of spatiotemporal coordinates. For example, in dispersion of scalars in laminar pipe flows\cite{Taylor1953}, averaging is performed in the pipe cross-section. In RANS modeling, the averaging dimensions are those over which the system is statistically homogeneous. In most cases, the temporal dimension is statistically stationary. In the absence of statistical stationarity, averaging is defined over many ensembles. A more advanced definition of $\overline{c}$, but outside of the scope of this report, utilizes filters or weighted averages. 

The problem is to determine the``macroscopic" operator, $\overline{\mathcal{L}}$, with an appropriate set of boundary conditions such that the solution to equation
\begin{equation}
\overline{\mathcal{L}}\overline{c}=0
\label{eq:macroscopic}
\end{equation}
matches exactly the $\overline{c}$ obtained from averaging the microscopic solution. Since the macroscopic space, has a lower dimension than the microscopic space, or better said, involves a significantly lower number of degrees of freedom, access to the macroscopic operator allows tremendous savings in computational cost. As we shall see, $\overline{\mathcal{L}}$ will depend on the statistics of the microscopic advective field $u_j$, $D_M$, the microscopic boundary conditions, and also the definition of averaging that converts $c$ to $\overline{c}$. 

The immediate difficulty with the above problem statement is that there are numerous macroscopic operators that satisfy the stated requirements. In other words, the answer to the problem is not unique. In what follows we will utilize additional physics-based constraints such that $\overline{\mathcal{L}}$ will be uniquely determined. 

A common method for seeking $\overline{\mathcal{L}}$, referred to as the Reynolds decomposition\cite{Reynolds1895},  is to apply the averaging operator directly to Equation~(\ref{eq:microscopic}) and analyze the resulting terms. For some terms in ${\mathcal{L}}$, averaging commutes with the operator in that term. For example, in Equation~(\ref{eq:advdiff}), ensemble averaging commutes with the time derivative and the diffusion operator,  but not with multiplication by $u_j$ in the advective term. Therefore, after averaging Equation~(\ref{eq:microscopic}), and noting that $\partial u_j/\partial x_j=0$, one can write
\begin{equation}
\label{eq:advdiffavg}
\frac{\partial \overline{c}}{\partial t}+\frac{\partial}{\partial x_j}\left(\overline{u_jc}\right)-\frac{\partial}{\partial x_j}\left(D_{M}\frac{\partial \overline{c}}{\partial x_j}\right)=0.
\end{equation}
If averaging had commuted with multiplication by $u_j$, for example in the case of a constant $u_j$, then one could have written the advective term in~(\ref{eq:advdiffavg}) as ${\partial\left(\overline{u_j}\ \overline{c}\right)}/{\partial x_j}$, and therefore the job of finding $\overline{\mathcal{L}}$ would have been complete. However, commutation is rarely permissible when the underlying microscopic flow field is non-uniform. With this line of thinking, the problem of finding the macroscopic operator is then focused on obtaining closures to $\overline{u_jc}$. Consistent with this mindset, we sometimes use the notation $\overline{\mathcal{L^\prime}}$ to refer to the portion of $\overline{\mathcal{L}}$ that cannot be closed due to commutation error. In other words, for the case of advection-diffusion problem one can write: 
\begin{equation}
\overline{\mathcal{L}}= \frac{\partial}{\partial t}+\overline{u_j}\frac{\partial}{\partial x_j}-\frac{\partial}{\partial x_j}\left(D_{M}\frac{\partial }{\partial x_j}\right) + \overline{\mathcal{L^\prime}}. 
\end{equation}
Use of this notation might mistakenly imply that $\overline{\mathcal{L^\prime}}$ is solely dependent on the microscopic flow velocity fluctuations, and independent of other processes such as molecular diffusion. We must remember that this is generally not the case; $\overline{\mathcal{L^\prime}}$ depends on all microscopic processes including those involving commuting operators. 
%For example, for the problem of dispersion by laminar flows $\overline{\mathcal{L^\prime}}$ is a strong function of the molecular diffusivity. 
% In contrast to the Reynolds decomposition method, our approach first derives the full macroscopic operator, $\overline{\mathcal{L}}$. But using the above equation, one can solve for the closure operator, $\overline{\mathcal{L^\prime}}$. 

When the microscopic advective process is limited to length scales much smaller than the macroscopic field,
% for example, a spatially periodic velocity profile with effective ``mean free path" much smaller than the domain size, 
the macroscopic operator  $\overline{\mathcal{L^\prime}}$ can be approximated by a diffusion operator\cite{Majda1999}. However, this condition is rarely met in turbulent flows. 

 Before proceeding to the next section, we introduce some additional terminology. We use $\Omega$ to refer to the microscopic phase space, and $\overline{\Omega}$ to refer to the macroscopic phase space. With this definition, $c\in\Omega$ and $\overline{c}\in\overline{\Omega}\subset\Omega$. In other words, the number of the degrees of freedom in $\overline{c}$ is (much) smaller than the those in $c$ owing to elimination of degrees of freedom by the averaging operator.  Averaging is a projection operator mapping  points (fields) from $\Omega$ to $\overline{\Omega}$. The microscopic operator, $\mathcal{L}$, maps points within $\Omega$, and the macroscopic operator maps points within $\overline{\Omega}$. While we refer to $\overline{\mathcal{L}}$ as the macroscopic operator, we use ``macroscopic closure operator" to refer to $\overline{\mathcal{L^\prime}}$ . In general, the macroscopic closure operator can be anisotropic, may involve non-constant coefficients, may involve spatial derivatives higher than second order, and may even be non-local.% We use ``standard eddy diffusivity" to refer to closure models that use the Boussinesq approximation. 
  
Next, we review, as a pedagogical example, the problem of dispersion by parallel flows. In Section~\ref{sec:simpleexample} we present an approximate solution to this problem consistent with the methodology of G. I. Taylor\cite{Taylor1953}. Then after introducing the macroscopic forcing method (MFM) in Section~\ref{sec:MFM}, we will revisit this problem and assess the performance of the macroscopic operators obtained using MFM and those obtained from Taylor's solution. 

\subsection{Simple Example: Dispersion by a Parallel Flow}
\label{sec:simpleexample}
A macroscopic model for dispersion of passive scalars by parallel flows was first introduced by G. I. Taylor\cite{Taylor1953}, where he considered the evolution of a passive scalar by a laminar pipe flow. Consistent with experimental observations, his analysis predicted that the long-term evolution of the cross-sectional averaged scalar field can be predicted by a diffusion equation to the leading order with an effective diffusivity quadratically dependent on the velocity magnitude and inversely proportional to the molecular diffusivity. Since then, this problem has been extensively revisited\cite{Aris1960,Ajdari2006,Barton1983,Vedel2012} with some studies offering corrections to the leading-order model\cite{Gill1970,Barton1983} including extensions to unsteady flows\cite{Chatwin77}, and non-parallel flows, such as those in porous media\cite{Koch1987,Frankel1989,Mauri1991}. We will show that MFM results are consistent with these corrections, while providing a systematic and robust computational approach for the determination of the macroscopic operators in more general settings including macroscopically inhomogeneous flows.  

%In this section, we focus on a simplified version of such flows, just to capture the key lessons and ingredients that we will later expand on in the broader context.
Without loss of key physical ingredients and lesson points, we consider a simplified parallel flow in a two-dimensional domain as depicted in Figure~\ref{fig:fig1}.a
\begin{equation}
u_1=U\cos\left(\frac{2\pi}{L_2}x_2\right), \hskip 0.4cm u_2=0,
\end{equation}
where $U$ is the velocity amplitude and $L_2$ is the domain length in the $x_2$-direction. We consider a microscopic transport equation given by
\begin{equation}
\frac{\partial c}{\partial t}+u_j\frac{\partial c}{\partial x_j}=D_M\frac{\partial^2c}{\partial x_1^2}+D_M\frac{\partial^2c}{\partial x_2^2}.
\label{eq:taylorpde}
\end{equation}
We consider the physical domain $-\infty < x_1 <+\infty $, and $0\le x_2<L_2$ with periodic boundary conditions in the $x_2$ direction. Here, we define the averaging operator
\begin{equation}
\overline{c}(x_1)= \frac{1}{L_2}\int_0^{L_2}c\left(x_1,x_2\right) \text{d} x_2.
\end{equation}
Our task is to determine the macroscopic differential operator acting only in the $x_1$ direction that describes the evolution of $\overline{c}$ without requiring a direct solution to (\ref{eq:taylorpde}). 

It is worth starting the analysis with a non-dimensionalization of coordinates. Scaling the spanwise length by $L_2/(2\pi)$, the streamwise length by $UL_2^2/(4\pi^2D_M)$, and time by $L_2^2/(4\pi^2D_M)$, results in the following dimensionless PDE:
\begin{equation}
\frac{\partial c}{\partial t}+\cos\left(x_2\right)\frac{\partial c}{\partial x_1}=\frac{\partial^2c}{\partial x_2^2}+\epsilon^2\frac{\partial^2c}{\partial x_1^2}, 
\label{eq:ndtaylor}
\end{equation}
defined over a domain $-\infty < x_1 <+\infty $, and $0\le x_2<2\pi$, with periodic boundary conditions in the $x_2$ direction. $\epsilon=2\pi D_M/(L_2U)$ is the only dimensionless parameter of the problem, representing an inverse Peclet number, Pe. Here, we introduce our mathematical tools and analyses considering only the regime of large Pe, and for a simplified version of the microscopic equation, in which we neglect the last term on the right hand side of (\ref{eq:ndtaylor}) with the intention of keeping this pedagogical toy problem brief. Later, we will present a more complex example in which finite $\epsilon$ effects are considered.

 Figure~\ref{fig:fig1}.c shows an example solution to this problem.  The macroscopic average of this solution, $\overline{c}$, is shown in Figure~\ref{fig:fig1}.d. The evolution equation for $\overline{c}$ can be derived by applying the averaging operator to Equation~(\ref{eq:ndtaylor}). In the averaging process, the first term on the right hand side of (\ref{eq:ndtaylor}) vanishes due to periodic boundary conditions. Considering the aforementioned simplification, the averaged equation can be written as
\begin{equation}
\frac{\partial\overline{c}}{\partial t} + \frac{\partial}{\partial x_1}\left(\overline{\cos(x_2)c}\right)=0.
\label{eq:ndtayloravg}
\end{equation}
The first term in~(\ref{eq:ndtayloravg}) is already closed as seen before, while the second term involves an unclosed product. Next, we derive an approximate closure to this equation following Taylor's approach. Although Taylor did not examine zero-mean harmonic parallel flows, we still refer to the solutions presented in this section as Taylor's solution.   

In the limit that the concentration field involves features with large streamwise length $l\gg 1$, often associated with long development time, it is possible to derive an approximate solution to Equation~(\ref{eq:ndtaylor}) via a perturbation approach. In this case, one can conclude that the time scale of mixing in the streamwise direction, $t_1~\sim O(l)$, is much longer than the time scale of mixing in the spanwise ($x_2$) direction via diffusion, $t_2\sim O(1)$. Therefore, concentration fields can be considered almost well-mixed in the spanwise direction (or more accurately said, ``rapidly  developed") while undergoing slower evolution by the streamwise flow. Introducing the decomposition $c(x_1,x_2)=\overline{c}(x_1) + c'(x_1,x_2)$,  fast mixing in the spanwise direction implies that $c'\ll\Delta \overline{c}$, where $\Delta\overline{c}$ represents variation of the averaged concentration in the streamwise direction. More fundamentally, rapid development in the spanwise direction, allows a quasi-steady approximation for the evolution of $c'$. By subtracting Equation~(\ref{eq:ndtayloravg}) from Equation~(\ref{eq:ndtaylor}) one can obtain an exact evolution equation for $c'$: 
\begin{equation}
\frac{\partial^2 c'}{\partial x_2^2}-\cos\left(x_2\right)\frac{\partial \overline{c}}{\partial x_1}=\frac{\partial c'}{\partial t}+\frac{\partial}{\partial x_1}\left[\cos\left(x_2\right)c'\right]',
\label{eq:taylorperturb}
\end{equation}
where the right-most prime ($^\prime$) on the square bracket implies deviation from the mean of the term inside the bracket. Given $c'\ll\Delta \overline{c}$, $\partial/\partial x_1\sim O(1/l) \ll 1$, and the quasi-steady assumption, the leading-order balance must be between the two terms on the left hand side. This results in a leading order solution for $c'$ as $c'^0=-\cos\left(x_2\right)\left({\partial \overline{c}}/{\partial x_1}\right)$. Substituting this expression into~(\ref{eq:ndtayloravg}) results in the following macroscopic equation for the evolution of $\overline{c}$
\begin{equation}
\frac{\partial \overline{c}}{\partial t}=\frac{1}{2}\frac{\partial^2 \overline{c}}{\partial x_1^2}.
\label{eq:taylormac}
\end{equation}
We refer to Equation~(\ref{eq:taylormac}) as Taylor's solution.According to this approximate solution, the macroscopic operator is $\overline{\mathcal{L}}\simeq \frac{\partial}{\partial t}-\frac{1}{2}\frac{\partial^2}{\partial x_1^2}$, while the second term alone represents the macroscopic closure operator, $\overline{\mathcal{L}^\prime}=-\frac{1}{2}\frac{\partial^2}{\partial x_1^2}$.

Dimensionally, this equation implies that the cross-sectional averaged concentration field experiences a macroscopic diffusivity equal to $U^2L_2^2/(8\pi^2D_M)$. The most notable outcome of this expression is the unintuitive inverse dependence on the molecular diffusivity, which is well explained in the literature and confirmed experimentally. 
%Although Taylor's work led to a tremendous practical impact, extension of his derivation to more general flows, such as turbulent flows, does not seem tractable.  Additional difficulty in pursuing such extensions is the limitation of large spanwise length assumption, which is not a valid assumption under general conditions. 

Before introducing our work, we briefly note that for the specific problem of parallel flows, improvements to Taylor's solution have been investigated\cite{Gill1970,Barton1983}. In what is presented above, the only approximation made in the solution to~(\ref{eq:taylorperturb}) was ignoring the right hand side terms. One may improve this approximation by considering a series expansion for $c'$ as $c'=\sum_{i=0}^{\infty}c'^i(x_1,x_2)$ in which $c'^i\sim O(\Delta \overline{c}/l^{i+1})$, and substitution in~(\ref{eq:taylorperturb}). Considering $\partial/\partial x_1\sim O(1/l)$ and $\partial/\partial t\sim O(1/l^2)$, as inferred from~(\ref{eq:taylormac}), and equating terms of similar order, one may obtain a recursive relation through which higher order solutions to $c'$ can be successively obtained by substitution of lower order solutions into the right hand side of~(\ref{eq:taylorperturb}). Substituting the improved $c'$ in~(\ref{eq:ndtayloravg}), results in an improved macroscopic model.  For example, the next correction results in the following macroscopic PDE:
\begin{equation}
\label{eq:taylormacimproved}
\frac{\partial}{\partial t}\left(\overline{c}+\frac{1}{2}\frac{\partial^2\overline{c}}{\partial x_1^2}\right)=\frac{1}{2}\frac{\partial^2\overline{c}}{\partial x_1^2}+\frac{1}{32}\frac{\partial^4\overline{c}}{\partial x_1^4}.
\end{equation}
Equation~(\ref{eq:taylormacimproved}) is a perturbative correction to Taylor's solution, with the macroscopic closure operator represented as $\overline{\mathcal{L}^\prime}=-\frac{1}{2}\frac{\partial^2}{\partial x_1^2}+\frac{1}{2}\frac{\partial^3}{\partial t\partial x_1^2}-\frac{1}{32}\frac{\partial ^4}{\partial x^4}$. The unfamiliar term, $\partial^3\overline{c}/\partial t\partial x_1^2$ can be simplified as $(1/2)\partial^4\overline{c}/\partial x_1^4 + O(1/l)^6$ as suggested by~(\ref{eq:taylormac}). Substituting this result in~(\ref{eq:taylormacimproved}) leads to a simpler macroscopic PDE, $\partial \overline{c}/\partial t=(1/2)\partial^2\overline{c}/\partial x_1^2-(7/32)\partial^4\overline{c}/\partial x_1^4$ with coefficient signs consistent with dissipation mechanism.  

By following the necessary algebraic steps one may realize that even in this simple setting, the correction procedure is cumbersome and quickly becomes analytically intractable. 

\section{The Macroscopic Forcing Method}
\label{sec:MFM}
We now resume the general problem by envisioning an arbitrary transport process described by a microscopic operator, $\mathcal{L}$. In discretized space we can write the microscopic equation as

\begin{equation}
    [\mathcal{L}][c]=0.
\end{equation}
This equation is identical to Equation~(\ref{eq:microscopic}), however we use brackets to denote matrices and vectors, reminding that the problem at hand is essentially a linear algebra problem. Our goal is to determine the macroscopic operator $[\overline{\mathcal{L}}]$ that acts directly on the macroscopic field $[\overline{c}]$. The answer is straightforward once one sees the problem from a linear algebra perspective: the macroscopic solution is simply projection of the microscopic solution to the mean space, which itself can be represented as a linear operation. Proper combination of the projection operator and the microscopic operator should therefore provide the macroscopic operator. We will discuss an approach based on this mindset later in Section~\ref{sec:LA}, however, we point that this direct approach would require determination of the inverse of the microscopic operator, which is prohibitively expensive for practical problems. We therefore, present the macroscopic forcing method as an alternative method for obtaining the same $[\overline{\mathcal{L}}]$ without requiring inversion of the microscopic operator.

%and seek to determine the macroscopic operator, $\overline{\mathcal{L}}$, such that the solution to Equation~(\ref{eq:macroscopic}) results in the same $\overline{c}$ as that obtained from averaging of the microscopic solution.
%While in Section~\ref{sec:MFMNS} we will generalize our method to the Navier-Stokes equation, for now we only consider advection-diffusion transport of scalar fields. 
Given the linearity of the microscopic operator, it is straightforward to deduce that the macroscopic operator, $\overline{\mathcal{L}}$, must also be linear. %It is however, difficult to obtain $\overline{\mathcal{L}}$ analytically. We therefore envision a computational procedure that can be used systematically to reveal this operator. By recognizing the fact that $\overline{\mathcal{L}}$ acting on $\overline{c}$ is analogous to matrix-vector multiplication, our task is to determine the elements of the matrix representing $\overline{\mathcal{L}}$. 
We can therefore write the macroscopic equation as 
\begin{equation}
\left[\overline{\mathcal{L}}\right]\left[\overline{c}\right]=0,
\label{eq:matrixzero}
\end{equation} 
In order to determine $\overline{\mathcal{L}}$, we examine how the macroscopic system above responds to forcing by setting up the input-output relation
\begin{equation}
\overline{\mathcal{L}}\overline{c}\left(s\right)=s,
\label{eq:matrix}
\end{equation} 
where $\overline{c}\left(s\right)$ denotes the output response due to the input forcing $s$. One can reveal columns of the matrix representing $\overline{\mathcal{L}}^{-1}$ by obtaining $\overline{c}$ in response to activation of different elements in $s$. By combining these columns, it is possible to construct both $\overline{\mathcal{L}}^{-1}$ and $\overline{\mathcal{L}}$. 
%In a well-posed procedure, the dimension of $s$ must match that of $\overline{c}$. Therefore, $s$ must belong to the macroscopic space, i.e., $\overline{s}=s$. One can reveal columns of the matrix representing $\overline{\mathcal{L}}^{-1}$ by numerically obtaining $\overline{c}$ in response to activation of different elements in $s$. By combining these columns, it is possible to construct $\overline{\mathcal{L}}^{-1}$ as well as $\overline{\mathcal{L}}$. %We will later discuss a more economical procedure, which we call IMFM, that allows approximate determination of $\overline{\mathcal{L}}$ without obtaining its inverse. 
%The macroscopic forcing method (MFM) is the procedure of obtaining $\overline{c}$ in response to a macroscopic forcing $s$ when the macroscopic operator is not explicitly available. 

Note that operations above are defined in the macroscopic space and thus $s\in\overline{\Omega}$, meaning that the forcing must be macroscopic ($s=\overline{s}$). However, in the absence of $\overline{\mathcal{L}}$, one can obtain $\overline{c}\left(s\right)$ by directly solving the microscopic equation, (\ref{eq:microscopic}), with the macroscopic forcing added to its right hand side,
\begin{equation}
\mathcal{L}\left[c\left(x_1,...,x_n,t\right)\right]=s.
\label{eq:MFM}
\end{equation}
MFM is the procedure of determining $\overline{\mathcal{L}}$ by obtaining $\overline{c}$ in response to different macroscopic forcing scenarios. With the description given above, we obtain $\overline{\mathcal{L}}$ while bypassing the expensive step of inverting $\overline{\mathcal{L}}$, however, the expense of the procedure still remains a critical concern. Brute force MFM, as discussed above, is still expensive, since it requires many DNS calculations. We will present in sections~\ref{sec:IMFM} and~\ref{sec:construct} methods to substantially reduce the cost of MFM owing to certain smoothness properties of $\overline{\mathcal{L}}$. % These consist of inverse MFM (IMFM) and the use of kernel moments. % We note here that this procedure is primarily useful for testing of canonical problems to generate insights about required macroscopic operators for transport phenomena. 

In the description above, we skipped discussion of initial condition and boundary conditions. One may naturally incorporate these by extending the discretized system~(\ref{eq:matrixzero}) to include both initial and boundary points in $[\overline{c}]$, and properly adjusting the operator  $[\overline{\mathcal{L}}]$, and the right-hand-side of~(\ref{eq:matrixzero}). In doing so, we implicitly assume that both initial condition and boundary conditions are macroscopic, i.e. their description in the macroscopic space is the same as that in the microscopic space. This constraint, however, does not limit the applicability of MFM to turbulence modeling since most practical problems are insensitive to initial conditions, and involve boundary conditions that are indeed macroscopic (e.g., boundary condition remains intact between Navier-Stokes and RANS descriptions). Keeping this in mind, we revisit the problem of dispersion by parallel flows by constraining the initial condition to be macroscopic, $c|_{t=0}=c\left(x_1\right)$, as shown in Figure~\ref{fig:fig1}b.   

\subsection{Revisiting Dispersion by a Parallel Flow}
\label{sec:taylorrevisit}
As a pedagogical example, we revisit the problem of dispersion by parallel flows. This time, we apply MFM to reveal the exact macroscopic operator. The equation to be solved is 
\begin{equation}
\frac{\partial c}{\partial t}+\cos\left(x_2\right)\frac{\partial c}{\partial x_1}=\frac{\partial^2c}{\partial x_2^2}+s(x_1,t).
\label{eq:MFMTaylor}
\end{equation}
Note that since the macroscopic space does not involve $x_2$, $s$ is chosen to depend only on $x_1$ and $t$. %According to MFM, we should find $\overline{c}\left(x_1,t\right)$ in response to different forcing scenarios, and then represent the matrix of response in terms of differential operators.
Selection of $s$ as a delta function in space and time will result in a $\overline{c}$ field representing the Green's function associated with the macroscopic operator. However, here, a more insightful practice would be to continue the analysis in Fourier space. As we shall see, an additional advantage of forcing in Fourier space is the smooth dependence of the macroscopic operator on the spatial wavenumber, $k$,  and temporal frequency, $\omega$, allowing its accurate representation with a coarse sampling of the forcing space. Therefore, we consider 
\begin{equation}
s(x_1,t)=\exp\left(i\omega t+ikx_1\right).
\label{eq:MFMTaylorForce}
\end{equation}
After obtaining the direct solution to Equation~(\ref{eq:MFMTaylor}) and averaging the result, we will find 
\begin{equation}
\overline{c}\left(x_1,t\right)=\overline{\widehat{c}}\exp\left(i\omega t+ikx_1\right).
\end{equation}
Given~(\ref{eq:matrix}), $\overline{\mathcal{L}}$ then can be expressed in Fourier space as 
\begin{equation}
\widehat{\overline{\mathcal{L}}}=1/\overline{\widehat{c}}\left(\omega,k\right).
\label{eq:lbaroperator}
\end{equation}
With this procedure in mind, we proceed to solve Equation~(\ref{eq:MFMTaylor}) assuming a forcing in the form of~(\ref{eq:MFMTaylorForce}). We assume a solution of the form 
$$
c=\widehat{c}(\omega,k, x_2)\exp\left(i\omega t+ikx_1\right).
$$
Substituting this expression into Equation~(\ref{eq:MFMTaylor}) results in
\begin{equation}
\left[i\omega-\frac{\partial^2}{\partial x_2^2}+ik\cos\left(x_2\right)\right]\widehat{c}\left(\omega,k;x_2\right)=1, 
\label{eq:c_0}
\end{equation}
 which is the Mathieu equation. However, to better establish intuition about the results and specifically ease interpretation between Fourier and physical spaces, we seek solutions that are asymptotically expressible in terms of polynomials or simple operations acting on polynomials. Therefore, instead of expressing the solution in terms of Mathieu function we take the following semi-analytical approach. For a given $k$ and $\omega$, one may expand $\widehat{c}$ as
\begin{equation}
\widehat{c}\left(\omega,k;x_2\right)=\sum_{n=0}^\infty a_n\left(\omega,k\right) \cos(nx_2).
\label{eq:series}
\end{equation}
Substituting this expansion into~(\ref{eq:c_0}) results in a system of linear equations for the $a_n$s. We present a solution to this system in Appendix~\ref{ap:solution}. The macroscopic operator in Fourier space, $\widehat{\overline{\mathcal{L}}}$, can be determined by substituting $\overline{\widehat{c}}=a_0(k,\omega)$ from this solution into~(\ref{eq:lbaroperator}). From~(\ref{eq:ndtayloravg}) we remember that the sole closed term in the full operator is the $\partial/\partial t$ term, and thus the macroscopic closure operator can be computed as
\begin{equation}
\widehat{\overline{\mathcal{L}^\prime}}=1/\overline{\widehat{c}}\left(\omega,k\right)-i\omega 
\end{equation}

\begin{figure}
%\vspace{4 mm}
\includegraphics[width=0.8\textwidth]{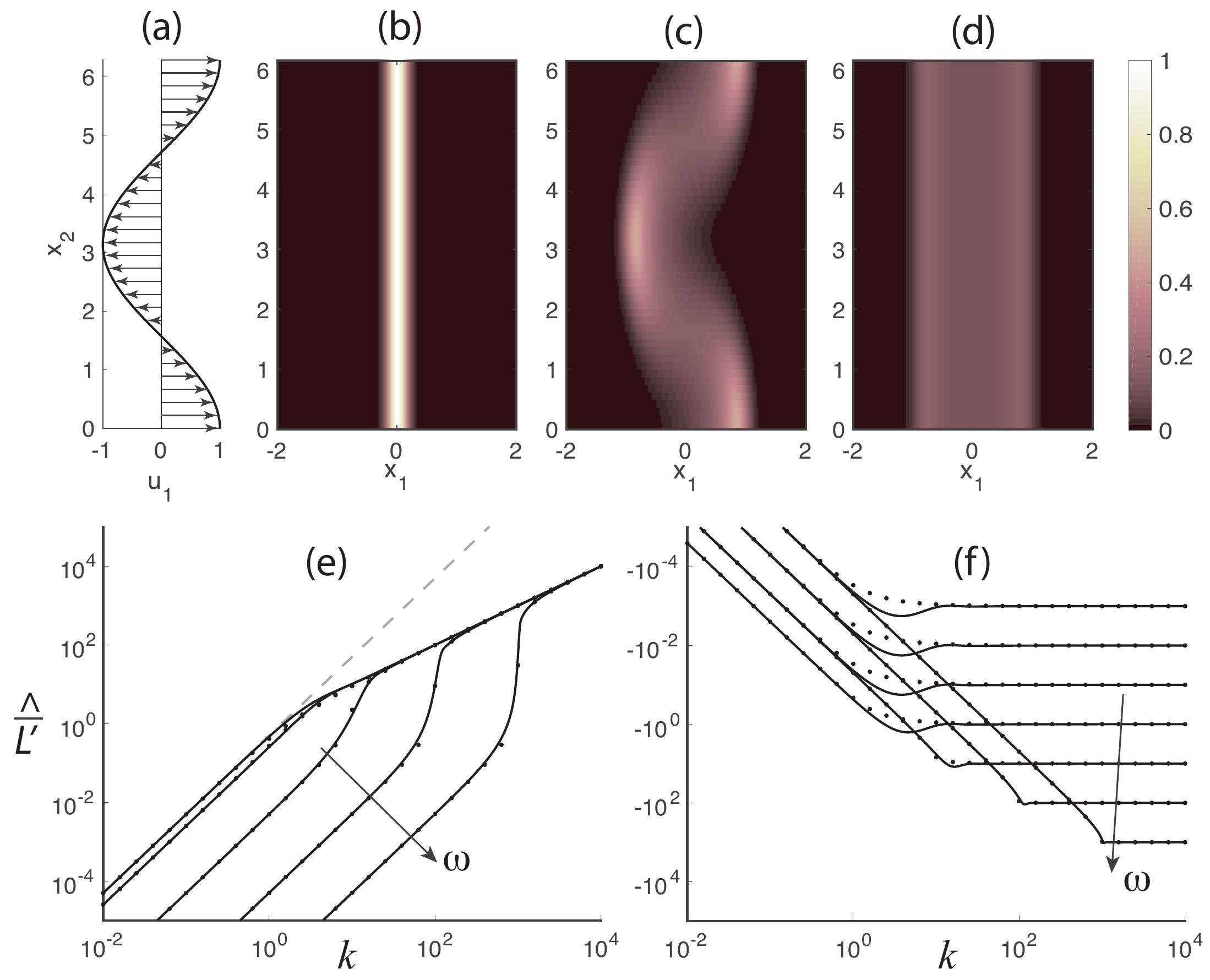}
\caption{Schematic of a parallel flow described by $u_1=\cos(x_2)$ (a), and an example initial condition for a scalar field to be dispersed by this flow (b) according to Equation~(\ref{eq:ndtaylor}) with $\epsilon=0$. Solution field at $t=1$ is shown in (c), and its projection to the macroscopic space is shown in (d). (e) and (f) respectively show the real and imaginary parts of the macroscopic closure operator,  $\widehat{\overline{\mathcal{L}^\prime}}$, versus $k$ for $\omega=10^{\{-3,-2,-1,0,1,2,3\}}$. Solid lines show the the exact values obtained through the procedure discussed in Appendix~\ref{ap:solution}, and the dotted symbols show the fitted operator of Equation~(\ref{eq:approxL}). In (e) the three first frequency cases are almost indistinguishable. The gray dashed line shows $\widehat{\overline{\mathcal{L}^\prime}}=k^2/2$ representing the diffusion operator derived from Taylor's solution in~(\ref{eq:taylormac}).}
\label{fig:fig1}
\end{figure}

Figures~\ref{fig:fig1}.e and \ref{fig:fig1}.f show the solution for $\widehat{\overline{\mathcal{L}^\prime}}\left(\omega,k\right)$ obtained from the above procedure. We first note that in the limit of small $k$ and $\omega$ the plotted solution matches with Taylor's solution given in (\ref{eq:taylormac}), which can be expressed in Fourier space as $\widehat{\overline{\mathcal{L}^\prime}}=0.5k^2$ (see Figure~\ref{fig:fig1}.e). Further examination indicates that higher-order terms in the Taylor series of this numerical solution also match the correction to the Taylor's solution given by (\ref{eq:taylormacimproved}), which is expressed in Fourier space as  $\widehat{\overline{\mathcal{L}^\prime}}=0.5k^2-0.5i\omega k^2-(1/32)k^4$. 

Unlike these perturbative approaches, the present MFM solution discloses the macroscopic closure operator over the entire spectrum of scales. As shown in Figures~\ref{fig:fig1}.e, for wavenumbers or frequencies of $~O(1)$ or higher, $\widehat{\overline{\mathcal{L}^\prime}}$ becomes significantly different from the operators from the perturbative models. In fact, even for $k$ slightly larger than one, the successive perturbative corrections lead to divergence from the true $\widehat{\overline{\mathcal{L}^\prime}}$ (not shown here). This is because perturbative corrections tend to approximate $\widehat{\overline{\mathcal{L}^\prime}}(k,\omega)$ in terms of a polynomial expansion where the order of the dominant term naturally increases as $k$ increases. In contrast, the true $\widehat{\overline{\mathcal{L}^\prime}}$ transitions from a second-order power-law $\widehat{\overline{\mathcal{L}^\prime}}\sim k^2$ to a lower-order power-law, $\widehat{\overline{\mathcal{L}^\prime}}\sim k^1$, as the magnitude of $k$ becomes larger as shown in the figure. 

\begin{figure}
%\vspace{4 mm}
\includegraphics[width=1.0\textwidth]{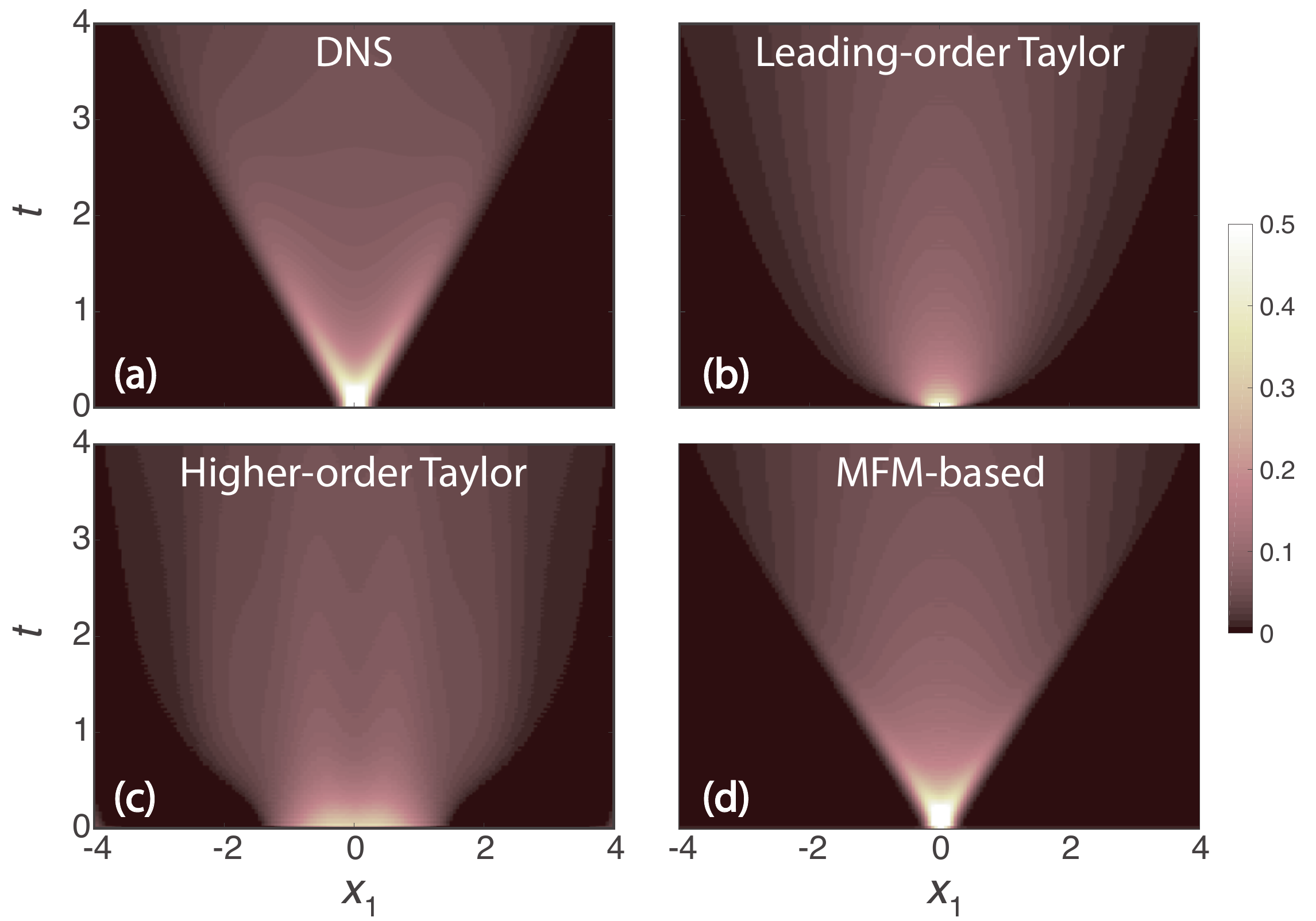}
\caption{Space-time evolution of $x_2$-averaged concentration field with an initial condition $c=\exp\left(-x_1^2/0.025\right)$ subject to the advection diffusion equation of~(\ref{eq:ndtaylor}): (a) DNS result, (b) prediction of Taylor's solution given by~(\ref{eq:taylormac}), (c) prediction of perturbative correction to Taylor's solution given by~(\ref{eq:taylormacimproved}), and (d) prediction of the fitted model obtained from MFM given by~(\ref{eq:approxLphys}) }
\label{fig:fig2}
\end{figure}

To demonstrate this point, we compare the performance of macroscopic closure models against a reference solution obtained from averaging of a DNS solution involving high wavenumbers. For the reference solution, we numerically solved the two-dimensional transport problem described by Equation~(\ref{eq:ndtaylor}). We considered $c(x_1,x_2,0)=\exp\left(-x_1^2/0.025\right)$ as an initial condition representing a thin zone of contaminant in a 2D space (see Figure~\ref{fig:fig1}.b). Using a second-order central difference code, and mesh resolutions $\Delta x_1=0.05$ and $\Delta x_2=0.04\pi$, the temporal evolution of the scalar field is obtained using the forth-order Runge-Kutta method with time step $\Delta t=0.002$. The resulting $\overline{c}\left(x_1,t\right)$ from 2D DNS is shown in Figure~\ref{fig:fig2}.a. Figure~\ref{fig:fig2}.b shows the Taylor's solution, given by Equation (\ref{eq:taylormac}) for the same setup. It is evident that the early-time dispersion is not captured properly. More remarkably, the perturbative correction to Taylor's solution, given by (\ref{eq:taylormacimproved}), predicts a worse behavior as shown in Figure~\ref{fig:fig2}.c. This lack of convergence has also been noted by others\cite{Frankel1989}, but it is insightful to revisit it here by examining the operators in Fourier space. 

It is needless to say that the full macroscopic closure operator, obtained from MFM, predicts a solution matching the mean of the DNS solution. While we have verified this claim numerically, this exercise is of little practical utility unless we can express the macroscopic closure operator in terms of a closed mathematical equation. In the absence of analytical solutions, we attempted to obtain analytical curve fits to the plots shown in Figure~\ref{fig:fig1}.e and~\ref{fig:fig1}.f. In doing so, we learned that capturing the true asymptotic behavior of the curves is more crucial than capturing the higher order terms in their Taylor series expansion. By examining the asymptotic behavior of $\widehat{\overline{\mathcal{L}^\prime}}$  in the limits of small and large $\omega$ and $k$, we obtained the following fitted expression for $\widehat{\overline{\mathcal{L}^\prime}}$
\begin{equation}
\widehat{\overline{\mathcal{L}^\prime}}\simeq\sqrt{\left(1+i\omega\right)^2+k^2} - \left(1+i\omega\right).
\label{eq:approxL}
\end{equation}
Specifically, Equation~(\ref{eq:approxL}) is exact in the limits of small or large $k$ or $\omega$ as shown in Figure~\ref{fig:fig1}.e and f. Before, developing an intuitive comprehension of this macroscopic closure model, we assess its performance against DNS as done for previous closure operators (see numerical details in Appendix~\ref{ap:numerical}). Figure~\ref{fig:fig2}.d shows the space-time evolution of $\overline{c}$ obtained from a numerical solution to the macroscopic equation with the closure operator as in (\ref{eq:approxL}). It is evident that this MFM-inspired closure operator offers remarkable improvement compared to Taylor's solution and its perturbative correction. 

Given that the expression in (\ref{eq:approxL}) is not a polynomial expression, one may wonder how the macroscopic operator looks in physical space. Based on Equation~(\ref{eq:approxL}), the macroscopic PDE in physical space can be obtained by replacing $k^2$ and $i\omega$ with their appropriate operator representation in physical space. These are respectively, $-\partial^2/\partial x_1^2$, and $\partial/\partial t$. After adding back the closed $\partial/\partial t$ term, one obtains the full macroscopic equation as:
\begin{equation}
\left[\sqrt{\left(\mathcal{I} + \frac{\partial}{\partial t}\right)^2-\frac{\partial^2}{\partial x_1^2}}-\mathcal{I}\right] \overline{c}\left(x_1,t\right)=s\left(x_1,t\right),
\label{eq:approxLphys}
\end{equation}
where $\mathcal{I}$ represents the identity operator. Equation~(\ref{eq:approxLphys}) represents an MFM-inspired macroscopic model for dispersion of scalars by the considered parallel flow. This mathematical outcome may seem surprising at first, because the resulting operator looks very different from those describing macroscopic dispersion, such as those in Taylor's solution and its perturbative corrections. This is one aspect of our study that we will continue to highlight repeatedly since the observed difference offers opportunity to improve closure models in use today. However, before moving forward with this mission, we build more intuition about this result by making the following observations: 
%Figure~\ref{fig:fig1} shows the fitted operator introduced in~(\ref{eq:approxL}) in comparison to the exact operator. The limits of large and small frequencies and wavenumbers are matched exactly, while there is satisfactory comparisons in the intermediate scales. It is possible to improve the remaining discrepancy by adding to the complexity of the fitted operator. We will revisit this later, but the current operator captures the key physical messages that we would like convey below: 

1- As a sanity check, we note that the resulting model in~(\ref{eq:approxL}) offers a spatially conservative operator. Given the form of the closure in~(\ref{eq:ndtayloravg}), the macoscopic operator must be divergence of a flux. Here the divergence is implicitly embedded in the expression, and can be verified by checking that the closure operator vanishes in the limit of $k=0$, noting that the intended root for the expression under the square root expands as $1+i\omega + O(k^2)$.
  
2- The resulting model involves a square root acting on an operator. The mathematical interpretation of this square root in Fourier space is straightforward: it implies algebraic square root acting on a number that depends on $k$ and $\omega$. In physical space, the mathematical interpretation can be eased if one thinks about how to implement this square root numerically: in the discrete space, operators are represented by matrices. The square root simply implies the square root of the matrix. We will build more physical intuition about the meaning of this square root in Section~\ref{sec:LA}. Use of fractional-order operators for closure models has been suggested before\cite{Chen2006,Epps2018}. However, we here directly obtain the operator from the governing equation, as opposed to inferring it from fitting solutions to experimental data of mean fields. Additionally, the obtained operator differs from common fractional-order closure operators, since the closure term is not solely $\partial^\alpha\overline{c}/\partial x_1^\alpha$,  with $\alpha$ being a non-integer constant. Neither do we obtain such behavior in any asymptotic limit. As we examine more cases and build more intuition, our initial excitement about fractional-order operators fades away and is replaced with a more general concept involving non-locality of operators as discussed in Section~\ref{sec:LA}. 

3- In the limit of small $k$ and $\omega$, Equation~(\ref{eq:approxL}) reduces to $\widehat{\overline{\mathcal{L}^\prime}}=k^2/2$. In other words, the macroscopic model in physical space is $\partial \overline{c}/\partial t=(1/2)\partial^2\overline{c}/\partial x_1^2$. This matches the Taylor's solution, presented in Equation~(\ref{eq:taylormac}), which is appropriate for long-term dispersion of contaminants by parallel flows. 

4- The key difference between the MFM-inspired closure operator and those in (\ref{eq:taylormac}) and (\ref{eq:taylormacimproved}), is in their character in the high-wavenumber limit, which has close connection to their performance in predicting early-time dispersion of contaminants as shown in Figure~\ref{fig:fig2}. The linear propagation of the contaminant front with time, shown in Figure~\ref{fig:fig2}.a, is due to the fact that the underlying mechanism of transport is advective. Taylor's solution, however, results in a diffusion equation, (\ref{eq:taylormac}), predicting contaminant propagation as $x_1\sim \sqrt{t}$, which implies an unbounded dispersion speed at early times proportional to $t^{-1/2}$. In contrast, the MFM-based model, (\ref{eq:approxLphys}), honors the physically expected dispersion speeds at both early and long times. This difference in physical space, can be traced in Fourier space by noting the character of $\widehat{\overline{\mathcal{L}^\prime}}$ in the high wavenumber limit. Unlike Taylor's solution, true $\widehat{\overline{\mathcal{L}^\prime}}$, and the proposed model in~(\ref{eq:approxL}) exhibit a high-wavenumber scaling as $\widehat{\overline{\mathcal{L}^\prime}}\sim k^1$. While there exist other models in the literature, such as the telegraph equation~\cite{Goldstein51}, that satisfy the property of finite dispersion speed, here the proposed MFM technique allows a systematic determination of the closure operator over all spatio-temporal scales. The obvious advantage is achieving quantitative accuracy, in addition to satisfying the expected qualitative trends. As a demonstration, in Appendix~\ref{ap:telegraph} we discuss a comparison of the solution obtained by the telegraph equation to our solution presented in Figure~\ref{fig:fig2}. Furthermore, in the steady limit ($\omega =0$), and in contrast to the telegraph equation, the true closure operator does not reduce to the standard diffusion operator, as we shall see next.

One motivation for examining the steady limit of the closure operator is that most practical turbulent flows are statistically time-stationary, and thus the RANS operator is a steady operator. Secondly, even in the context of semi-parallel laminar flows, sometimes the steady macroscopic operator is of interest. Though in such cases, instead of an initial value problem, one is interested in a boundary value problem.  In some cases, Taylor's solution is not satisfactory as one is interested in solutions beyond the small wavenumber limit\cite{Yaroshchuk2011,Rubinstein2013}. 

\subsection{Steady Limit of Dispersion by a Parallel Flow}
We examine the steady limit of the same problem discussed in Section~\ref{sec:taylorrevisit}. Given the homogeneity of this problem in the $x_1$ direction, we are not yet examining a boundary value problem. We will consider statistically inhomogeneous problems with boundary conditions in Section~\ref{sec:MFMIH}.

\begin{figure}
%\vspace{4 mm}
\includegraphics[width=0.5\textwidth]{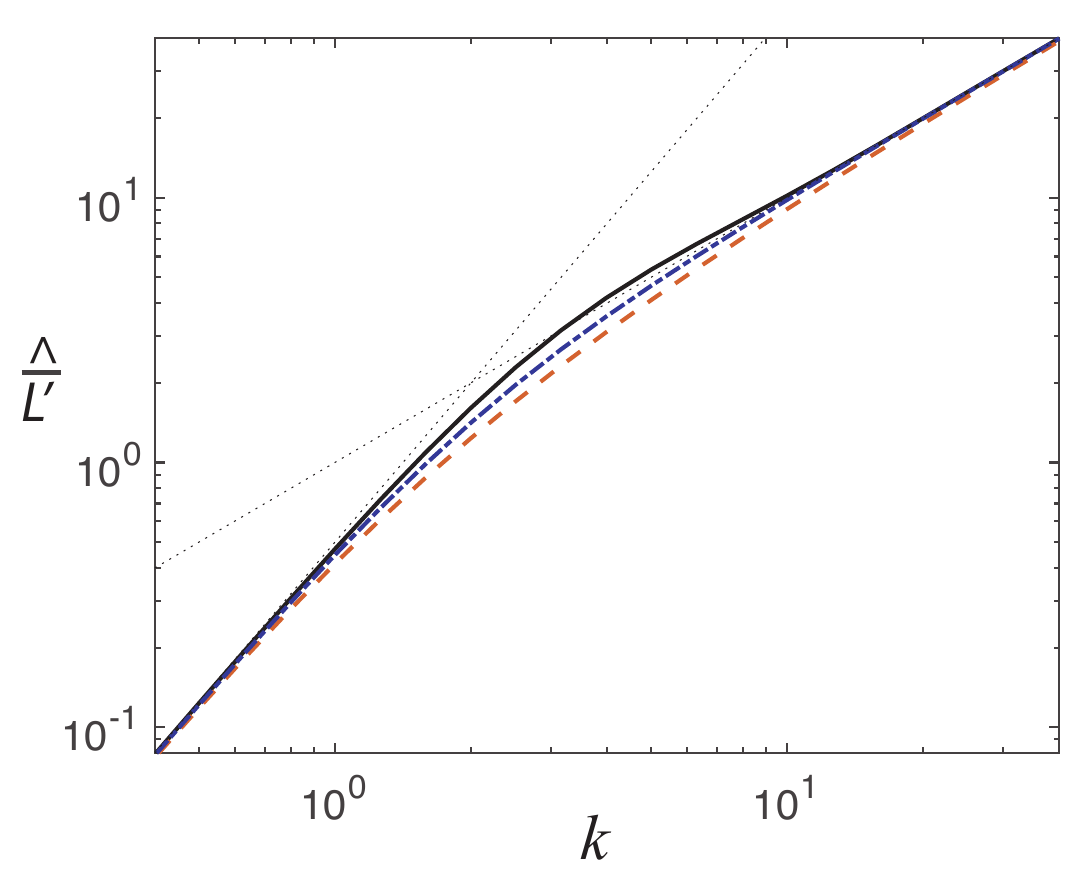}
\caption{Macroscopic closure operator,  $\widehat{\overline{\mathcal{L}^\prime}}$, versus $k$ for the steady limit, $\omega=0$, of the microscopic equation~(\ref{eq:ndtaylor}). The solid line shows the exact operator obtained from the MFM procedure discussed in Appendix~\ref{ap:solution}; the dashed line is $\widehat{\overline{\mathcal{L}^\prime}}=\sqrt{1+k^2}-1$; the dashed-dotted line shows $\widehat{\overline{\mathcal{L}^\prime}}=k^2/\sqrt{4+k^2}$. The thin dotted lines show the asymptotic limits of $0.5k^2$, and $k$, respectively in the small and large wavenumbers. The imaginary part of all solutions is zero.}
\label{fig:fig3}
\end{figure}

Figure~\ref{fig:fig3} shows the macroscopic closure operator for the steady limit ($\omega=0$) in wavenumber space. In the small wavenumber limit, where one may assume separation of scales between mean fields and underlying flow scale, the $k^2$ scaling is recovered. In the large wavenumber limit, as discussed earlier, a linear $k^1$ scaling prevails, even in the steady limit. The transition wavenumber between these two scalings is $k=2$, which is obtained from the intersection of (dotted) asymptotic lines in Figure~\ref{fig:fig3}. Converting back to dimensional units, this transition wavenumber is $8\pi^2D_M/(UL_2^2)$. The inverse of this scale, $UL_2^2/(8\pi^2D_M)$, plays a role analogous to the ``mean free path" in kinetic theory of gases: it is a length that is scaled by the product of contaminant's streamwise characteristic velocity, $U$, and the time scale $L_2^2/(8\pi^2D_M)$ for contaminants to diffusively cross a distance of order $L_2/2$ and switching the sign of their streamwise advective motion, hence completing a ``Brownian walk". Following Prandtl\cite{Prandtl1925}, we will refer to this length as the ``mixing length" instead of a mean free path.  

With this observation, the analogy with molecular to continuum macroscopic models becomes more clear. Taylor's diffusion model, (\ref{eq:taylormac}), suffices for macroscopic scales much larger than the mixing length, and thus analogue Knudsen number should be defined based on the mixing length. Furthermore, the macroscopic diffusion coefficient is equal to the product of the mixing length (mean free path), $UL_2^2/(8\pi^2D_M)$  and the microscopic velocity $U$. 

For macroscopic scales smaller than the mixing length, the boundedness of dispersion speed requires the macroscopic operator to change character to a $k^1$ scaling.  However, unlike the advection operator, which also scales as $k^1$, the nature of $\widehat{\overline{\mathcal{L}^\prime}}$ is dissipative and not advective. In other words, the pre-factor to this proportionality is real and positive, not imaginary. Therefore, a first derivative operator $\overline{\mathcal{L}^\prime}\sim \partial/\partial x_1$ does not capture the correct trends. This is why we used a square root to capture the expected asymptotic limit as $\overline{\mathcal{L}^\prime}\sim \sqrt{\partial^2/\partial x_1^2}$. However, this should be viewed only as a quick remedy; in Section~\ref{sec:LA} we present a generalized operator form after building more intuition about the problem.

%While the earlier discussion in Section~\ref{sec:taylorrevisit} explained this trend based on boundedness of dispersion speed, which is inherently an unsteady concept, we highlight here that $k^1$ scaling holds even for the steady limit.  

% Examination of the plotted data in the low $k$ limit suggests that the corrected leading-order behavior is $\widehat{\overline{\mathcal{L}^\prime}}=(1/2)k^2-(1/32)k^4$. This is consistent with the steady limit of the perturbative correction to Taylor's solution given by~(\ref{eq:taylormacimproved}). But this correction quickly diverges from the expected behavior at $k\sim O(1)$, and in practice only works in the low $k$ limit where the leading term alone is a good approximation to $\widehat{\overline{\mathcal{L}^\prime}}$. 

The fitted model presented in Equation~(\ref{eq:approxL}) reduces to $\widehat{\overline{\mathcal{L}^\prime}}=\sqrt{1+k^2}-1$ in the steady limit. In this limit, however, we were able to obtain a better fit (see Figure \ref{fig:fig3} dash-dotted),
\begin{equation}
\widehat{\overline{\mathcal{L}^\prime}}=\frac{0.5k^2}{\sqrt{1+0.25k^2}}.
\end{equation}
With this model the macroscopic closure operator in physical space can be written as
 \begin{equation}
 \overline{\mathcal{L}^\prime}=-\frac{\partial}{\partial x_1}\left[ \left(\frac{0.5}{\sqrt{\mathcal{I}-\frac{1}{4}\frac{\partial^2}{\partial x_1^2}}}\right) \frac{\partial}{\partial x_1}\right],
 \end{equation}
which is a divergence of a flux as expected. The flux itself is written as a macroscopic ``diffusivity" times the gradient of the mean field, $\mathcal{D}\partial/\partial x_1$, which is also a familiar form. We emphasize that the macroscopic diffusivity is a non-local operator,
\begin{equation}
\label{eq:deddy}
\mathcal{D}=\frac{D}{\sqrt{\mathcal{I}-l^2\frac{\partial^2}{\partial x_1^2} }},
\end{equation}
where $D=0.5$ is the diffusion coefficient from Taylor's analysis, and $l=0.5$ is the mixing length. In the low wavenumber limit, the second derivative in the denominator is negligible, and thus the macroscopic diffusivity is a number $\mathcal{D}=D=0.5$. Conversely, in the high-wavenumber limit, macroscopic diffusivity vanishes inversely proportional to $k$.  We remind that the eddy diffusivity operator introduced in ~(\ref{eq:deddy}) is non-singular. The denominator of this operator is diagonally dominant and invertible.

 Although the parallel flow problem considered here is a simplified example, the learning lessons presented have broad implications both regarding the method applied and regarding the obtained closure operator. For example, in \cite{Shirian2019} we show application of MFM to homogeneous isotropic turbulence, yet revealing a closure operator form identical to~(\ref{eq:deddy}).

Analysis in Fourier space limits the applicability of the presented method so far to problems in which the macroscopic space is spatially homogeneous. In such cases it is appropriate to utilize harmonic functions to construct a basis representing the solution space. In the next sections we extend MFM by departing from this simplified limit by considering inhomogeneous flows and domains involving wall boundary conditions. We demonstrate that an alternative solution-space basis in terms of polynomial functions provide a cost-manageable approach to MFM analysis.

\section{MFM for Inhomogeneous Flows}
\label{sec:MFMIH}
So far we have demonstrated how MFM can be applied to flows with homogeneous directions. In these cases, given the smoothness of $\overline{\mathcal{L}^\prime}$ when transformed to Fourier space, we were able to save on the number of required MFM simulations. For non-homogeneous problems, however, formulation in Fourier space is inappropriate. Nevertheless, the methodology presented in Section~\ref{sec:MFM} is extendable to non-homogeneous problems if one applies MFM in physical space. In what follows, we will first adopt this brute force approach to develop intuition about inhomogeneous macroscopic operators while ignoring the computational expense of MFM. In the following sections we will remedy the high expense of this brute force MFM, by introducing a more computationally feasible MFM, which focuses on computation of moments of eddy diffusivity kernels.

\subsection{Linear Algebra Interpretation}
\label{sec:LA}
In order to generate some mathematical insights, it is worth developing an understanding of MFM from a linear algebra point of view. The starting point is the microscopic equation, 
\begin{equation}
\label{eq:microLA}
\left[\mathcal{L}\right]\left[c\right]=\left[s\right],
\end{equation}
where $\left[\mathcal{L}\right]$ represents the discretized linear microscopic operator, $\left[c\right]$ is a vector representing the microscopic field, and $\left[s\right]$ is a vector representing the forcing. Square brackets indicate a matrix or vector involving a discrete set of numbers.% We refer to matrix and vector elements by subscripts after the brackets. 
One may represent the generalized momentum transport equation in a similar fashion by considering the field vector to represent momentum and pressure fields and the operator matrix to represent the discretized form of~(\ref{eq:gns}) and~(\ref{eq:gns2}) where the boundary conditions can be embedded in $\left[\mathcal{L}\right]$. 

Next, we define the averaging operator as
\begin{equation}
\label{eq:averaging}
\left[\overline{c}\right]=\left[P\right]\left[c\right],
\end{equation}
where $\left[ P \right]$ is the non-square project matrix with number of rows and columns respectively equal to the dimension of $\overline{\Omega}$ and $\Omega$. %When a uniform mesh is used for discretization, $\left[ P \right]_{ij}$ is $1/n$ if $\left[ c \right]_j$ contributes to $\left[ \overline{c} \right]_i$; otherwise $\left[P\right]_{ij}$ is zero. $n$ is the number of mesh points in the dimension where the averaging is performed. 
Given that the forcing is macroscopic, we have $\overline{s}=s$ in continuum space. In discrete space, however, $\left[\overline{s}\right]$ and $\left[s\right]$ have different dimensions. Therefore, before applying MFM, one needs to extend (or interpolate) $\left[\overline{s}\right]$ to the microscopic space 
\begin{equation}
\label{eq:averagings}
\left[s\right]=\left[E\right]\left[\overline{s}\right],
\end{equation}
where  $\left[E\right]$ is the non-square extension matrix with number of rows and columns respectively equal to the dimension of $\Omega$ and $\overline{\Omega}$. $\left[E\right]$ and $\left[P\right]$ satisfy the relation $\left[P\right]\left[E\right]=\mathcal{I}$. 

With these definitions at hand, we can obtain the macroscopic operator in terms of the defined matrices above. Combining equations~(\ref{eq:microLA}), (\ref{eq:averaging}) and (\ref{eq:averagings}), one can write 
\begin{equation}
\label{eq:bruteMFM}
\left[\overline{c}\right]=\left[P\right]\left[\mathcal{L}\right]^{-1}\left[E\right]\left[\overline{s}\right].
\end{equation}

From the definition of macroscopic operator in~(\ref{eq:matrix}), we have that $\left[\overline{c}\right]=\left[\overline{\mathcal{L}}\right]^{-1}\left[\overline{s}\right]$. Comparing this with~(\ref{eq:bruteMFM}) we conclude that
\begin{equation}
\label{eq:MFMLA}
\left[\overline{\mathcal{L}}\right]=\left\{\left[P\right]\left[\mathcal{L}\right]^{-1}\left[E\right]\right\}^{-1}.
\end{equation}
Equation~(\ref{eq:MFMLA}) is a foundational equation in the sense that it represents the macroscopic operator defined directly in terms of the microscopic operator and the definition of average which determines the projection and extension operators. Unlike MFM, this linear-algebra-based approach does not explicitly involve the force field $s$. However, computation of $\overline{\mathcal{L}}$ from this approach can be more expensive due to the cost of direct computation of $\left[\mathcal{L}\right]^{-1}$.  MFM is a remedy technique that avoids costly inversion of the microscopic operator by utilizing explicit forcing in order to reveal the macroscopic operator, $\overline{\mathcal{L}}$. Nevertheless, brute force MFM is still highly expensive.

\begin{figure}
%\vspace{4 mm}
\includegraphics[width=0.4\textwidth]{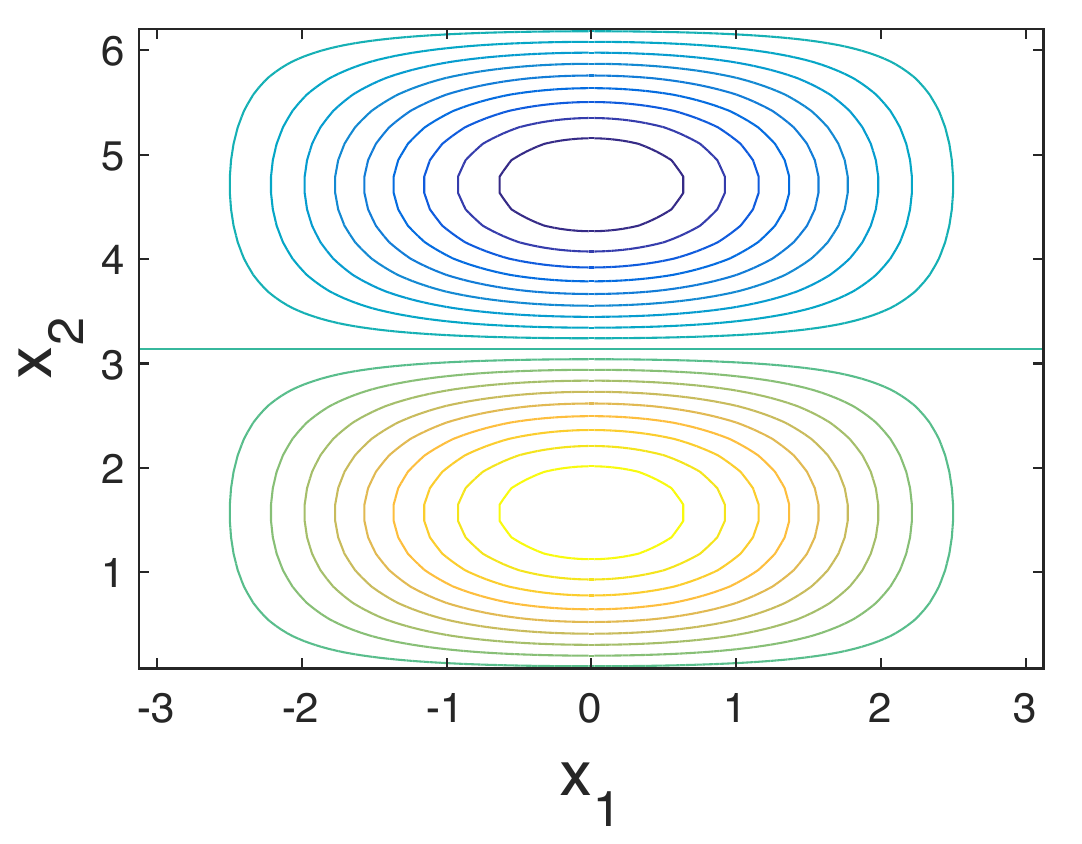}
\caption{Flow streamlines in a 2D channel. Velocity field is given by Equation~(\ref{eq:fig4}). }
\label{fig:fig4}
\end{figure}

Before remedying the cost issue, we develop more insights about macroscopic operators, by examining inhomogeneous mixing of a passive scalar using the described linear-algebra method based on Equation~(\ref{eq:MFMLA}). We consider a 2D domain representing a channel with the left and right walls at $x_1=\pm\pi$ with a Dirichlet condition $c=0$, and the top and bottom walls at $x_2=0,2\pi$ with a no-flux condition $\partial c/\partial x_2=0$. The concentration field $c\left(x,y\right)$ is governed by the steady advection diffusion equation
\begin{equation}
u_1\frac{\partial c}{\partial x_1} + u_2\frac{\partial c}{\partial x_2}=0.05\frac{\partial^2c}{\partial x_1^2} +\frac{\partial^2c}{\partial x_2^2}
\end{equation}
One may interpret the unequal diffusivities in the two directions to be an outcome of directional nondimensionalization as seen in Section~\ref{sec:simpleexample}, i.e., $\epsilon^2=0.05$. For this example, we consider an incompressible flow satisfying the no-penetration conditions on all walls described by (see Figure~\ref{fig:fig4})
\begin{equation}
\label{eq:fig4}
u_1=\left[1+\cos\left(x_1\right)\right]\cos\left(x_2\right), u_2=\sin\left(x_1\right)\sin\left(x_2\right).
\end{equation}

Discretization of the governing PDE is performed by utilizing the second-order central difference scheme on a uniform staggered mesh using $N_1$ and $N_2$ interior grid points in the $x_1$ and $x_2$ directions, respectively. The advective fluxes on the faces are computed using a second-order interpolation and then multiplication by the  divergence-free normal velocity at the face centers. Fluxes on the left and right boundaries are computed using the interior scheme after extrapolating the concentration field to ghost points half a grid size outside of the domain using a second-order extrapolation scheme that uses knowledge of the concentration field at the boundary and the two adjacent interior cells. To ensure a well-posed system, the number of degrees of freedom associated with the forcing term must be equal to those associated with the concentration field itself. Therefore, the forcing term is also defined at the cell centers corresponding to the same locations as the interior concentration fields. 

\begin{figure}
\includegraphics[width=0.8\textwidth]{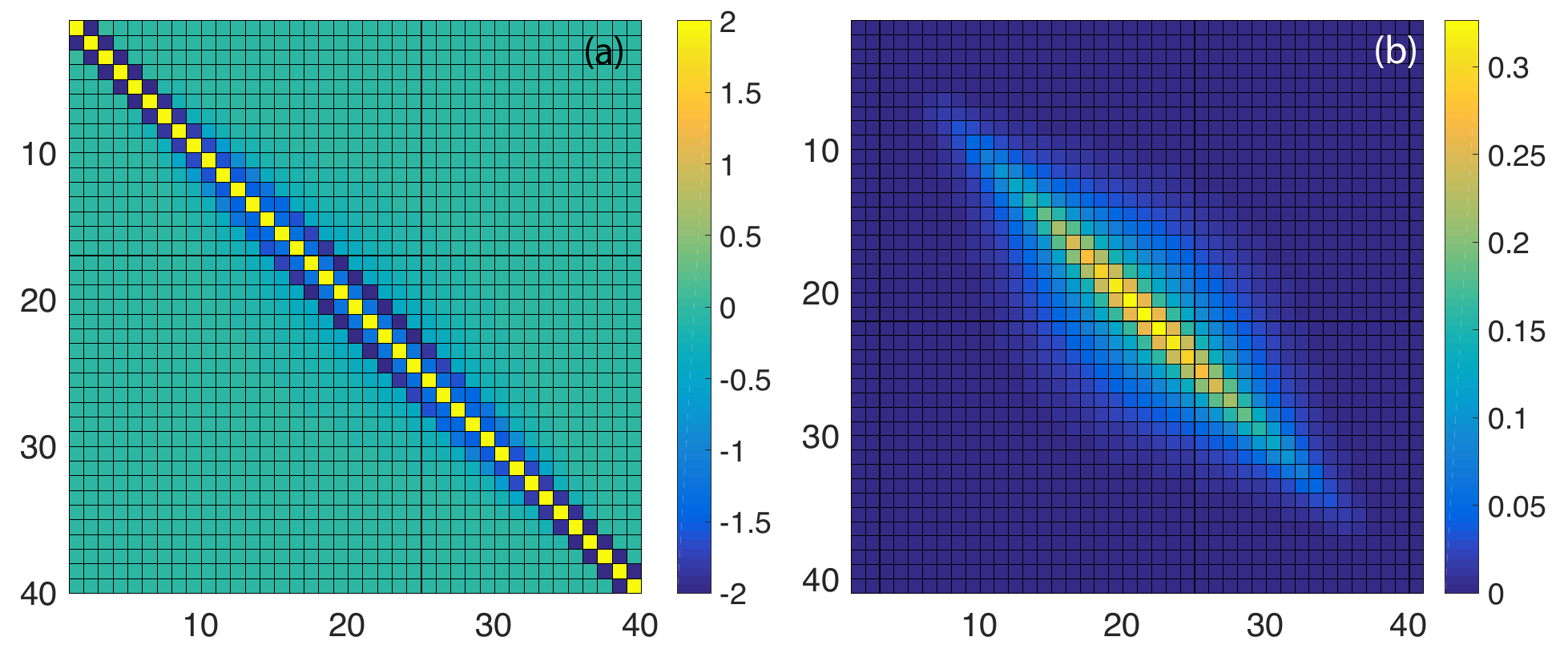}
\caption{(a) $\left[\overline{\mathcal{L}}\right]$ for the example problem described in Section~\ref{sec:LA} using $N_1=40$ and $N_2=10$. The color bar is intentionally truncated at $-2$ and $2$.  (b) the eddy diffusivity operator, $\left[\mathcal{D}\right]$, for the same problem.}
\label{fig:fig5}
\end{figure}

For this problem we define $\overline{c}$ as the average of the scalar field in the $x_2$ direction. Considering $N_1=40$ and $N_2=10$, Figure~\ref{fig:fig5}.a shows the matrix associated with $\left[\overline{\mathcal{L}}\right]$ obtained from implementation of Equation~(\ref{eq:MFMLA}) for this problem. One can see qualitative similarity between this matrix and the standard second-order diffusion operator. %This qualitative similarity indicates that eddy diffusivity is a reasonable qualitative model describing the macroscopic behavior. 
We next further investigate this numerical macroscopic operator to infer a continuum operator representing the eddy diffusivity as an operator, in a manner similar to Equation~(\ref{eq:deddy}).

\subsection{Eddy diffusivity as a non-local integration kernel}
 We confirmed computationally that $\left[\overline{\mathcal{L}}\right]$ can be written as the product of three matrices:
\begin{equation}
\left[\overline{\mathcal{L}}\right]=-\left[\partial/\partial x_1\right]\left[\mathcal{D}+D_M\mathcal{I}\right]\left[\partial/\partial x_1\right],
\end{equation}
where $D_M=0.05$ is the dimensionless molecular diffusivity in the $x_1$ direction. The right-most $\left[\partial/\partial x_1\right]$ is a gradient operator utilizing the aforementioned ghost point scheme, and the left-most $\left[\partial/\partial x_1\right]$ is a divergence operator acting on the fluxes defined on the cell faces. The middle operator is the total diffusivity, and $\left[\mathcal{D}\right]$ is the eddy-diffusivity matrix whose entries are shown in Figure~\ref{fig:fig5}.b. %Common eddy diffusivity models used in practice would have predicted a diagonal $\left[\mathcal{D}\right]$ with zeros for the non-diagonal entries. 
The non-locality of this operator is quantified by the off-diagonal entries in Figure~\ref{fig:fig5}.b. It physically implies that the macroscopic flux at a location is not just dependent on the macroscopic gradients at the same location. Instead, one needs to combine the macroscopic gradients from a neighborhood. The spatial extent of this neighborhood is scaled by the mixing length.

%%%ALI READ UP TO HERE JAN 20th 
% To DO: 

% X change standard to local and leading-order when appropriate 
% X distinction between brute-force and linear algebra 
% Search for keywords: X Taylor, X local, X leading, X standard, X diffusion, X eddy, X diffusivity, operator, macroscopic, microscopic make sure they are used correctly 
% Go after Spalart corrections/comments

% shall we have "local eddy diffusivity" 

\begin{figure}
\includegraphics[width=0.8\textwidth]{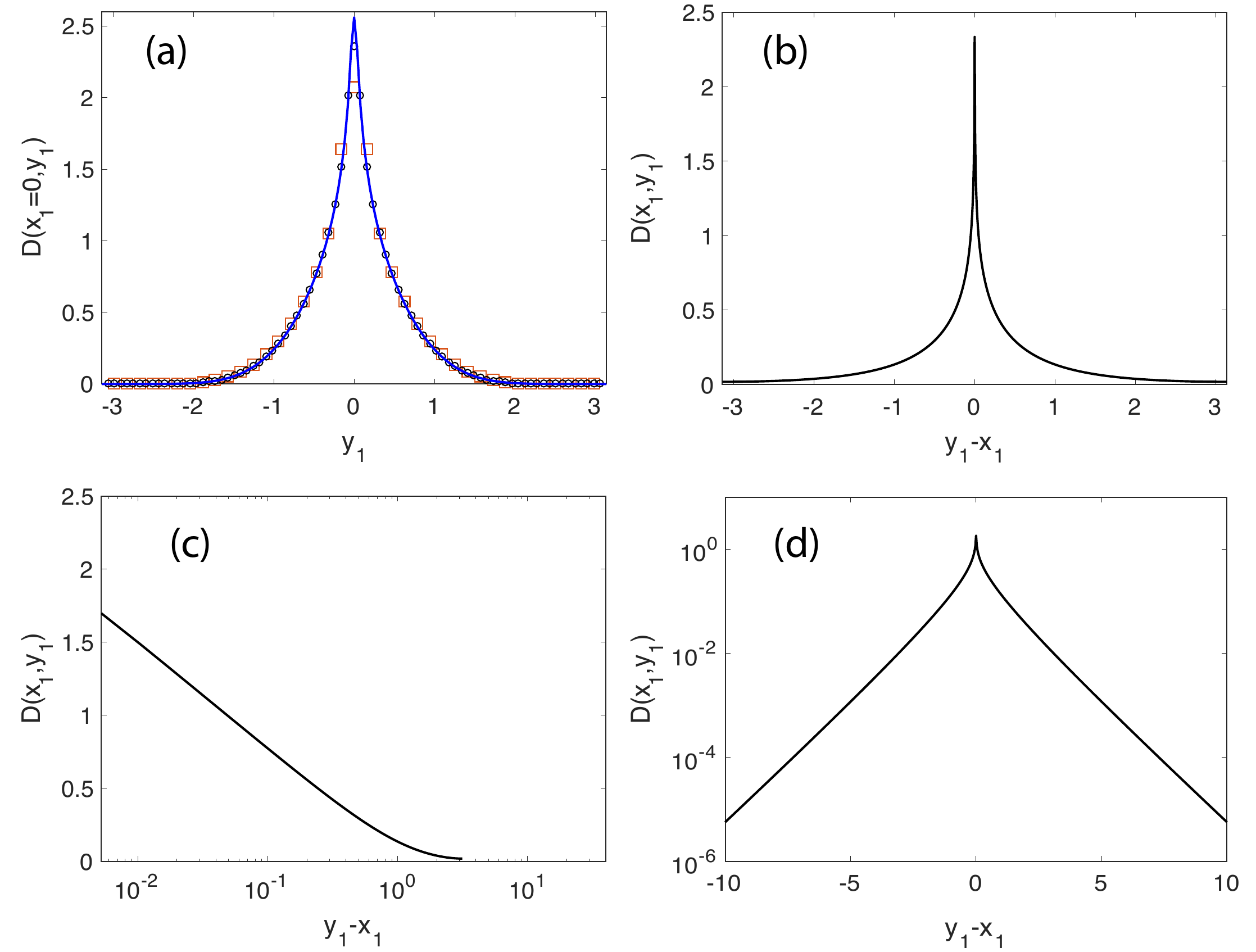}
\caption{(a) Numerically computed kernel at $x_1=0$ associated with $\mathcal{D}$ obtained from linear-algebra-based method for the described inhomogeneous problem. Different plots represent mesh refinement: squares $(N_1=40,N_2=10)$, circles $(N_1=80,N_2=20)$, solid line $(N_1=160,N_2=40)$. (b) eddy diffusivity associated with the homogeneous operator $\mathcal{D}=1/\sqrt{1-\partial^2/\partial x_1^2}$, also plotted in semi-log scales (c) and (d).}
\label{fig:fig6}
\end{figure}

Translating this result back to continuum space, we conclude that eddy diffusivity is a non-local operator that outputs the (negative of the) closure flux in terms of a weighted integral acting on the mean gradient of the transported quantity. In this case $\mathcal{D}$ is the integration kernel, representing how mean gradient in one location can cause mean fluxes in another location in the domain:
\begin{equation}
\label{eq:1dkernel}
-\overline{u_1^\prime c^\prime}(x_1)=\int_{y_1}\mathcal{D}\left({x_1},{y_1}\right)\frac{\partial \overline{c}}{\partial x_1}|_{y_1}{dy_1}.
\end{equation}
Extending this to multi-dimensions, eddy diffusivity is described by a tensorial kernel, $\mathcal{D}_{ji}\left({\bf x},{\bf y}\right)$, that expresses the macroscopic fluxes in terms of the macroscopic gradients as 
\begin{equation}
\label{eq:fluxkernel}
-\overline{u_j^\prime c^\prime}({\bf x})=\int_{\bf y}\mathcal{D}_{ji}\left({\bf x},{\bf y}\right)\frac{\partial \overline{c}}{\partial x_i}|_{\bf y}{\bf dy}.
\end{equation}
%We predict that a similar result holds for the case of transported vector fields with the difference that the eddy-diffusivity operator will be described by a kernel involving a fourth-order tensor as
%\begin{equation}
%\label{eq:kernelvect}
%-\overline{u_j^\prime v_i^\prime}({\bf x})= \int_{\bf y}\mathcal{D}_{jilk}\left({\bf x},{\bf y}\right)\frac{\partial \overline{v_k}}{\partial x_l}|_{\bf y}{\bf dy}.
%\end{equation}
For the numerical example considered above, we show that the implied kernel from our discrete solution, $\mathcal{D}(x_{1,i},y_{1,j})=\left[\mathcal{D}\right]_{ij}/{\Delta x_1}$, converges by mesh refinement (here, the problem is 1D, and  subscripts after the bracket imply discrete mesh points). As an example, we show in Figure~\ref{fig:fig6}.a the plot of $\mathcal{D}(x_1=0,y_1)$ versus $y_1$ obtained after successive mesh refinements. 

Below we report the major observations, and some speculations regarding their validity for more general settings:

1- For cases with zero normal velocity near a boundary and with boundary conditions that remain invariant under averaging, the eddy-diffusivity kernel vanishes near the boundary (see Figure~\ref{fig:fig5}.b). Therefore, the macroscopic operator reduces to the molecular diffusivity operator near a boundary. This implies there is no need for additional boundary conditions when scaling up to the macroscopic fields. 

2- In the present example, the macroscopic space was 1D and the eddy-diffusivity involved a positive kernel. For multidimensional systems the kernel is tensorial and entries can be positive or negative. It will be interesting to examine however, whether the diagonal entries of the tensor are positive-valued, and whether the tensor is positive definite.  

3- In the present example, the computed $[\mathcal{D}]$ was symmetric to the machine precision. This implies that if a mean gradient at $x_{1A}$ result in a mean flux at $x_{1B}$, the same mean gradient applied to $x_{1B}$ will result in same mean flux at $x_{1A}$. In other words, there is reciprocity in eddy diffusivity operator. It will be interesting to investigate under what conditions this reciprocity holds, and whether it is generalizable to cases with mean advection as well as cases with tensorial eddy diffusivity. 

4- When the macroscopic dimensions are statistically homogeneous the eddy-diffusivity kernel can be simplified as  $\mathcal{D}({\bf x},{\bf y})=\mathcal{D}({\bf x}-{\bf y})$, and thus convolution in physical space will result in multiplication in the Fourier space. The specific fitted form for the eddy-diffusivity operator identified earlier as $\mathcal{D}={D}/\sqrt{\mathcal{I}-l^2\partial^2/\partial x_1^2}$ (see Equations~\ref{eq:deddy}) results in a qualitatively similar kernel as that obtained for the inhomogeneous problem (see Figure~\ref{fig:fig6}.b). Specifically, the kernel is positive for all values of $y_1$ and the kernel width scales with the mixing length, $l$, as alluded to earlier. 

5- In the limit that the mixing length (i.e. kernel width) is much smaller than the macroscopic length, the eddy-diffusivity kernel can be safely approximated by a Dirac delta function. Mathematically, this can be justified since the macroscopic gradient can be assumed constant over the kernel width and taken out of the integral. In other words, the integration can be approximated as multiplication by the area under the kernel. This is what Taylor's solution, presented in Section~\ref{sec:simpleexample}, achieves. Smallness of the mixing length is also one of the key conditions necessary for the validity of the Boussinesq approximation, but it is rarely met in realistic turbulent flows. However, even in this limit, the eddy diffusivity is not necessarily an isotropic tensor further challenging the validity of the Boussinesq approximation. %MFM is still useful in quantification of full tensorial properties associated with mixing by the underlying flow. 

6- In earlier sections we saw universally in all examples that in the limit of large wavenumber, the macroscopic closure operator is proportional to $k^1$, implying that $\widehat{\mathcal{D}}\sim k^{-1}$. Inverse transforming this relation to physical space implies that the eddy-diffusivity kernel, $\mathcal{D}({\bf x},{\bf y})$, must be proportional to $\log\left(1/|{\bf x}-{\bf y}|\right)$ for small separation distances in macroscopic 1D settings as shown in Figure~\ref{fig:fig6}.c. Consistently $\mathcal{D}({\bf x},{\bf y})\sim 1/|{\bf x}-{\bf y}|$ in macroscopic 2D and $\mathcal{D}({\bf x},{\bf y})\sim 1/|{\bf x}-{\bf y}|^2$ in macroscopically 3D settings. These are all singular but integrable kernels. %This singularity is also observed in the last test case as seen in Figure~\ref{fig:fig6}, given the values on the axis has an increase with mesh refinement, but remaining converged everywhere else. 

7- In earlier sections we also observed that the eddy-diffusivity operator is a smooth curve in Fourier space. We here assume an infinitely differentiable curve in $k$-space. This implies that the kernel associated with the eddy diffusivity, $\mathcal{D}({\bf x},{\bf y})$, must vanish exponentially in the limit of large $|{\bf x}-{\bf y}|$ as shown in Figure~\ref{fig:fig6}.d.

\subsection{Inverse MFM for determination of kernel moments}
\label{sec:IMFM}
The insights gained from examination of the full macroscopic operator in Section~\ref{sec:LA} allows development of a more economical method for approximate determination of the macroscopic operator. To do this, we first introduce the inverse macroscopic forcing method (IMFM). The prior macroscopic forcing method, which should be called forward MFM (FMFM), determines $\overline{c}$ in response to a specific macroscopic forcing. In contrast, IMFM obtains the macroscopic forcing required for sustaining a pre-specified $\overline{c}$. For problems in which the microscopic PDE involves a time stepping process, IMFM is straightforward. Given the PDE
\begin{equation}
\frac{\partial c}{\partial t}= f(c, \nabla c, t, ...) + s,
\end{equation}
one simply can take the average of this equation to arrive at 
\begin{equation}
\frac{\partial \overline{c}}{\partial t}= \overline{f} + s.
\end{equation}
By expanding the time discrete form of the left hand side term, and substituting $\overline{c}\left(t+\Delta t\right)$ with the pre-specified $\overline{c}$ one obtains an explicit expression for $s=\overline{s}$ which can be substituted in the equation. For example, for the forward Euler time-stepping method $s=\left[\overline{c}\left(t+\Delta t\right)-\overline{c}\left(t\right)\right]/\Delta t-\overline{f}\left(t\right)$. Substituting this expression in the microscopic PDE constrains $\overline{c}$ to be the pre-specified value, but the non-macroscopic modes in $c$ are allowed to evolve. 

For problems with a steady microscopic PDE, this treatment can be used in conjunction with artificial time stepping, where the true $s$ is evaluated in the long term limit. For turbulent flow problems that are statistically time stationary, the microscopic PDE is still time dependent. In such cases, $s$ is time-independent but is needed instantaneously. One may run multiple ensembles of the simulation and use the ensemble average at each instant. In the limit of infinite ensembles, one will obtain a time stationary $s$. However, one may run simulations using a finite number of ensembles and use the time average of the computed $s$ as an improved estimate of the true $s$. We observed this approach to be satisfactory and converge quickly with number of ensembles in our MFM analysis of turbulent channel flows (not presented here). While running simultaneous ensembles may increase computational cost per time step, it does not affect the total cost since convergence can be achieved over a proportionally lower number of time-steps. 

Lastly we note that when performing IMFM, the initial condition and boundary conditions must be adjusted to be consistent with the input $\overline{c}$. For example, while the left-right boundary conditions in the example discussed in Section~\ref{sec:LA} remain of Dirichlet type, the specified values of $c$ at the boundary must be consistent with those of $\overline{c}$ evaluated at the boundary.  

The introduced IMFM methodology can be used as another brute force method to reveal columns of matrices associated with closure operators, or even in Fourier space for homogeneous problems. The immediate advantage of IMFM is that one can utilize it to directly probe a desired closure term without first encountering the full macroscopic operator. Our intention here however, is to utilize IMFM to develop a cost-effective way of determining the closure operator. 

With the IMFM methodology described above one may use a small number of simulations to determine the low-order spatial moments of the eddy-diffusivity kernel, $\mathcal{D}({\bf x},{\bf y})$. For example, for the macroscopically 1D problem discussed in Section~\ref{sec:LA}, by selecting $\overline{c}=x_1$ and performing one IMFM DNS, and post-processing the $\overline{u_1^\prime c^\prime}$ data, one finds the zeroth moment of the kernel as (substitute $\partial \overline{c}/\partial x_1 =1$ into Equation~\ref{eq:1dkernel}),
\begin{equation}
D^0(x_1)=\int_{y_1}\mathcal{D}\left(x_1,y_1\right)dy_1=-\overline{u_1^\prime c^\prime}|_{\overline{c}=x_1}.
\end{equation}
In other words, with only one MFM calculation we obtain the area under the the eddy diffusivity kernel curve versus $y_1$ for all $x_1$ locations (see the plot in Figure~\ref{fig:fig6}.a as an example). By specifying successively higher order polynomial expressions for $\overline{c}$ versus $x_1$ it is possible to determine the higher spatial moments of the kernel. The first moment of the eddy-diffusivity kernel, defined below, can be computed by selecting $\overline{c}=x_1^2/2$, performing IMFM, and post processing the resulting $\overline{u_1^\prime c^\prime}$. Combining with the results of the zeroth moment, and analytical manipulation of Equation(~\ref{eq:1dkernel}) one can write 
\begin{equation}
D^1(x_1)=\int_{y_1}\left(y_1-x_1\right)\mathcal{D}\left(x_1,y_1\right)dy_1=-\left(\overline{u_1^\prime c^\prime}|_{\overline{c}=x_1^2/2}-x_1\overline{u_1^\prime c^\prime}|_{\overline{c}=x_1}\right), 
\end{equation}
where the last term on the right hand side can be substituted using $D^0$. Likewise, the second moment of $\mathcal{D}$ can be computed as
\begin{equation}
D^2(x_1)=\int_{y_1}\frac{1}{2}\left(y_1-x_1\right)^2\mathcal{D}\left(x_1,y_1\right)dy_1=-\left(\overline{u_1^\prime c^\prime}|_{\overline{c}=x_1^3/6}-x_1\overline{u_1^\prime c^\prime}|_{\overline{c}=x_1^2/2}+\left(x_1^2/2\right)\overline{u_1^\prime c^\prime}|_{\overline{c}=x_1}\right). 
\end{equation}
One may generalize this recursive methodology to compute higher moments of $\mathcal{D}$. Determination of the leading-order moments requires few IMFM DNSs, and thus it is much less costly than the brute force MFM. Given that the kernel is smooth away from the origin ($y_1=x_1$) it is possible to accurately reconstruct it from information of the leading order moments. 

%As we shall see, $D^0$ is the coefficient of the leading-order macroscopic closure operator, which is the usual diffusion operator. In the limit that the mixing length is much smaller than the macroscopic scale, this will be the dominant operator. For multi-dimensional {\color{blue} macroscopic} problems, $D^0$ is replaced by a tensor due to its anisotropy, i.e. the Boussinesq law is not followed. 

%To quantify the non-local effects, higher order moments must be computed. 

%The additional advantage of IMFM, either in the context of moments or even when used as a brute force method, is that it allows direct probing of the desired unclosed terms in the macroscopic operator. For example, in this case, one may directly determine the operator associated with the unclosed flux $\overline{u_1^\prime c^\prime}$, without requiring first obtaining the full macroscopic operator $\overline{\mathcal{L}}$ (as done in FMFM). Aside from cost saving, closure of individual fluxes, as opposed to the full unclosed term, has the advantage of guaranteeing conservativeness of the macroscopic model. For more complicated microscopic systems that may involve multiple unclosed terms, IMFM allows probing of different unclosed terms independently, and thus provides more insightful guidelines for modeling and understanding of the underlying dynamics. Note that in IMFM, the forcing field $s$ is not explicitly used when examining the macroscopic operator. 

Next, we discuss how the macroscopic operators can be constructed with reasonable accuracy from the knowledge of these moments. 

\subsection{Construction of the Macroscopic Operators from Kernel Moments}
\label{sec:moments}
Having moments of $\mathcal{D}$ at hand, the task is to approximate the integration kernel in~(\ref{eq:fluxkernel}). We first consider a non-convergent, but conceptually insightful, approximation. 

\subsubsection{Approximation of kernel integral using Taylor series}
The integral in Equation~(\ref{eq:fluxkernel}) can be rewritten by expanding $\partial\overline{c}/\partial x_i$ in terms of its spatial Taylor series around ${\bf y}={\bf x}$. Substitution and utilization of the defined moments leads to representations of $\overline{u_j^\prime c^\prime}$ in terms of a series involving spatial derivatives of $\overline{c}$ with the computed moments as pre-factors to each term. For example, for the test case considered in Section~\ref{sec:LA}, the unclosed flux, and thus the macroscopic closure operator, can be modeled as  
\begin{equation}
\label{eq:expansion}
-\overline{u_1^\prime c^\prime}\left(x_1\right)=\left[D^0\left(x_1\right)+D^1\left(x_1\right)\frac{\partial}{\partial x_1}+D^2\left(x_1\right)\frac{\partial^2}{\partial x_1^2}+ ...\right]\frac{\partial \overline{c}}{\partial x_1}.
\end{equation}
The leading term in this expansion, $D^0$, is the local approximation to the eddy-diffusivity operator; it approximates the kernel with a Dirac delta function with matching integral. %This term is the dominant term when the eddy length or ``mean free path" is small compared to the macroscopic scale. %Truncation of the eddy diffusivity operator at this level implies that the turbulent flux can be approximated based on the local gradient. %For multidimensional problems, $D^0$ is a tensor, which in general can be anisotropic. Therefore, even a leading-order truncation can involve non-Boussinesq effects, which can be quantified for the first time using IMFM. 
Higher order terms in~(\ref{eq:expansion}) will be present only for a non-local eddy diffusivity where the spatial moments of the kernel are finite. These terms can become significant when the macroscopic scale becomes small enough to be comparable to the kernel's characteristic width, i.e., the mixing length. 

\begin{figure}
\includegraphics[width=0.8\textwidth]{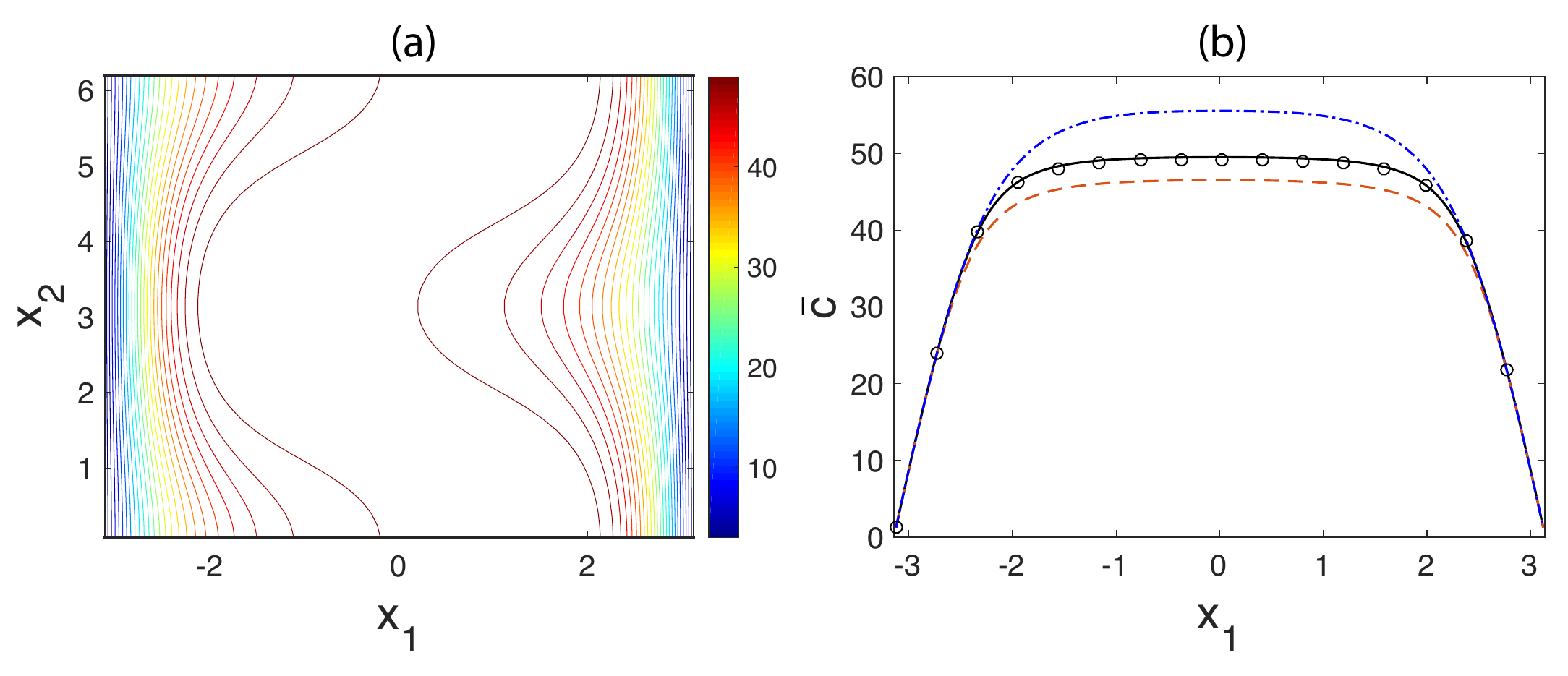}
\caption{(a) contour plots representing concentration field computed from 2D DNS of Equation~(\ref{eq:DNS}). (b) concentration averaged in the $x_2$ direction: solid, DNS; dashed, eddy diffusivity represented by Equation~~(\ref{eq:expansion}) but including only $D^0$; dash-dotted, eddy diffusivity represented by Equation~(\ref{eq:expansion}) including $D^0$ and $D^1$ terms; circle, eddy diffusivity obtained from the kernel reconstruction method.}
\label{fig:fig7}
\end{figure}

One might expect that a truncated expansion based on a few moments beyond $D^0$ could result in considerable model improvement. Such expansion would still lead to a well-posed differential equation, since coefficients (to higher derivatives) vanish near the boundary as the kernel itself vanishes, and thus a model based on this expansion does not impose requirement of additional boundary conditions. However, the expansion presented in Equation~(\ref{eq:expansion}) is not useful in practice since addition of higher moments does not lead to convergence of the solution. For example, considering the test case discussed in Section~\ref{sec:LA}, we use DNS of the following microscopic system to assess performance of macroscopic models,
\begin{equation}
\label{eq:DNS}
u_1\frac{\partial c}{\partial x_1} + u_2\frac{\partial c}{\partial x_2}=0.05\frac{\partial^2c}{\partial x_1^2} +\frac{\partial^2c}{\partial x_2^2}+1. 
\end{equation}
Figure~\ref{fig:fig7}.a shows the 2D DNS results and Figure~\ref{fig:fig7}.b shows the macroscopic fields. A macroscopic model that retains only the leading operator, $D^0$, performs a reasonable job in predicting the qualitative features of the mean field, while still lacking quantitatively. When expansion~(\ref{eq:expansion}) is used, addition of the new terms did not lead to prediction improvement despite relatively smooth mean fields. 

This observation is consistent with the issue seen in Section~\ref{sec:taylorrevisit} when we examined homogeneous parallel flow systems. In fact the macroscopic closure operator in the Taylor's solution represented the zeroth moment of the macroscopic diffusivity kernel for the considered flow while its perturbative correction was indeed an analytical derivation of the higher moments in the expansion of a convolution integral (with the exception that in the analytical derivation in Section~\ref{sec:taylorrevisit} we considered the unsteady problem and thus captured both spatial and temporal moments). As previously shown in Figure~\ref{fig:fig2}.c, addition of terms in the moment expansion does not result in improved macroscopic prediction. 

It turns out, the expansion in~(\ref{eq:expansion}) is analogous to the Kramers-Moyal expansion\cite{Van1981} previously formulated for the statistical description of stochastic processes associated with Brownian dynamics. Lack of convergence of this expansion has been previously investigated as described by the Pawula theorem. Likewise, in these applications, while the leading moment promisingly captures a significant portion of the macroscopic behavior, addition of higher moments do not improve quantitative prediction. We are aware of one previous work\cite{Mauri1991} reporting analogy to the Kramers-Moyal expansion in the continuum context in their analytical study of laminar transport of reacting scalars in porous media. %Compared to this analytical work, the novelty of IMFM is in that it provides a systematic framework for quantitative determination of moments of eddy-diffusivity kernel, including systems with macroscopic spatial inhomogeneity, and its extension to the Navier-Stokes system.  

\subsubsection{Approximate kernel reconstruction}
\label{sec:construct}
Although Equation~(\ref{eq:expansion}) is not a useful expansion in practice, the computed moments from IMFM are still valuable in the sense that one can utilize them to construct converging macroscopic operators. We next show an example of such construction. Our approach has been guided by the fact that all higher-order operators beyond $D^0$ are inherently non-local. We therefore construct approximate integration kernels that match the moments obtained from MFM. The details of the procedure for the example problem in Figure~\ref{fig:fig7} are as follows. 
%and 2) the non-locality is expressible by a positively valued kernel. For homogeneous systems, the fitted operator of Equation~(\ref{eq:hiteddy}) satisfies both of these properties. For inhomogeneous systems, we developed a kernel reconstruction method as follows. 

We first used IMFM to compute the first three moments of $\mathcal{D}$, i.e., $D^0, D^1, D^2$. We then constructed positively valued kernels that matched the computed moments at each $x_1$. One approach would be to assume known kernel profiles, and tune their parameters to match the low-order moments (see Appendix~\ref{ap:kernel}). We then used the constructed kernel to setup the macroscopic equation as, 
$$
\frac{\partial}{\partial x_1}\left[\int_{y_1}\mathcal{D}\left({x_1},{y_1}\right)\left(\frac{\partial\overline{c}}{\partial x_1}\right)|_{y_1}d{y_1} + 0.05 \frac{\partial\overline{c}}{\partial x_1}\right]+1=0.
$$
This macroscopic equation can be solved numerically after discretization. For practical problems involving 3D integrals, one may need to design iterative techniques for fast approximate calculation of the integral. Figure~\ref{fig:fig7}.b shows the macroscopic solution obtained through the reconstructed kernel method. Compared to prediction of the model that retains only the leading operator, $D^0$, the solution from the reconstructed kernel, not only shows convergence, but also significant quantitative improvement.

In the next section we discuss how the concepts developed in sections~\ref{sec:MFM} and~\ref{sec:MFMIH} can be extended to determine closure operators for momentum transport. 

\section{Macroscopic Forcing Method for the Navier-Stokes Equation}
\label{sec:MFMNS}
Extension of MFM to the Navier-Stokes equation (NS) requires discussion of two remaining issues. First is the fact that the microscopic solution is often time dependent, and turbulent. This issue can be treated by using ensemble averages. As an alternative, given that in many practical scenarios the statistical fields are time-stationary, it is sufficient to employ MFM for obtaining the steady macroscopic operator by considering time-independent forcing terms, while averaging is performed over time. The spatial inhomogeneity of such problems results in macroscopic operators that have space-dependent coefficients. Statistical anisotropy would lead to macroscopic operators of tensorial form. 

The second issue, which is the more challenging one, is the nonlinearity of the Navier-Stokes equation as a microscopic model. We resolve this issue by introducing a generalized equation as follows. 

Consider a velocity field, $u_i$, which is the solution to the incompressible Navier-Stokes equation
\begin{equation}
\label{eq:ns}
\frac{\partial u_i}{\partial t} + \frac{\partial u_j u_i}{\partial x_j}= - \frac{\partial p}{\partial x_i} + \nu \frac{\partial^2 u_i}{\partial x_j \partial x_j} + r_i,
\end{equation}
\begin{equation}
\label{eq:ns2}
\frac{\partial u_j}{\partial x_j}=0,
\end{equation} 
where $p$ is the pressure field normalized by fluid density, and $\nu$ is the kinematic viscosity, and $r_i$ is the known body force, which in most practical cases is either zero, or macroscopic, i.e., $r_i=\overline{r_i}$. For example, in turbulent channel flows $r_i$ represents the imposed mean pressure gradient, which is a unity vector field in the streamwise direction when reported in units of shear velocity and channel half height. $r_i$ may also represent the known inhomogeneous terms associated with the boundary conditions for discretized problems. 

Next, we define the Generalized Momentum Transport (GMT) equation for a given $u_i$ as
\begin{equation}
\label{eq:gns}
\frac{\partial v_i}{\partial t} + \frac{\partial u_j v_i}{\partial x_j}= - \frac{\partial q}{\partial x_i} + \nu \frac{\partial^2 v_i}{\partial x_j \partial x_j} + s_i,
\end{equation}
\begin{equation}
\label{eq:gns2}
\frac{\partial v_j}{\partial x_j}=0, 
\end{equation}
where $q$ is the generalized pressure to ensure incompressibility. The boundary conditions are the same type as those for $u_i$, but the forcing $s_i$ is allowed to be different from $r_i$. Equation~(\ref{eq:gns}), subject to constraint~(\ref{eq:gns2}), describes a linear system for $v_i$, and thus its macroscopic form can be obtained using MFM. In this case, ${\bf s}$ must be selected from $\overline{\Omega}$ and thus ${\bf s}=\overline{\bf s}$. Let us assume that this reduced form can be written as $\overline{\mathcal{L}}\left(\overline{\bf v}\right)=-\nabla \overline{q} + {\bf s}$ subject to the macroscopic incompressibility constraint, $\nabla.\overline{{\bf v}}=0$. The question is whether the resulting operator can produce the correct RANS solution. In other words, we should show that
\begin{equation}
\overline{\mathcal{L}}\left(\overline{\bf u}\right)=-\nabla \overline{p} + {\bf r},
\end{equation}
where ${\bf r}$ is the same body force as in~(\ref{eq:ns}), and subject to the same boundary conditions as those for the Navier-Stokes system. To show this condition, it is sufficient to conclude that when Equation~(\ref{eq:gns}) is supplemented with the same body force and boundary conditions as those in the Navier-Stokes, the resulting microscopic solution, ${\bf v}$ will have the same mean as the mean of Navier-Stokes solution, i.e., $\overline{\bf v}=\overline{\bf u}$. 

It turns out, an even stronger condition holds: when the macroscopic body force of GMT is matched with that of NS, not only will the mean solution of GMT match that of NS, but the instantaneous microscopic solution of GMT also merges (exponentially) to that of NS. This condition may seem unintuitive at first, since NS often admits chaotic solutions whose long term behavior is highly sensitive to the initial condition. How could the solution to such systems merge to identical fields regardless of the initial condition? The answer lies in the linearity of GMT. A quick qualitative proof can be made by introducing the vector field ${\bf w}={\bf v}-{\bf u}$. The evolution equation for ${\bf w}$ can be obtained by subtracting NS from GMT, 
\begin{equation}
\label{eq:gnsw}
\frac{\partial w_i}{\partial t} + \frac{\partial u_j w_i}{\partial x_j}= - \frac{\partial \phi}{\partial x_i} + \nu \frac{\partial^2 w_i}{\partial x_j \partial x_j}
\end{equation}
\begin{equation}
\label{eq:gnsw2}
\frac{\partial w_j}{\partial x_j}=0, 
\end{equation}
where $\phi=q-p$ and the body force is identically zero, given ${\bf s}={\bf r}$. Additionally, since the boundary conditions of NS and GMT match, this equation is subject to homogeneous boundary conditions. Without the presence of any body force or energy injection from the boundaries, the energy in the field ${\bf w}$ shall decline over time. One may show this formally by deriving an evolution equation for the total kinetic energy of this difference field defined as $K=(1/2)\int\int\int w_iw_i dx_1dx_2dx_3$. Contracting (\ref{eq:gnsw}) with ${\bf w}$ and integrating in space results in
\begin{equation}
\frac{\partial K}{\partial t}=-\nu \int\int\int\left[ \frac{\partial w_i}{\partial x_j}\frac{\partial w_i}{\partial x_j}\right] dx_1dx_2dx_3,
\end{equation}
where integration by parts in conjunction with the homogeneous boundary conditions for ${\bf w}$ are used to arrive at the above equation. The only long term solution to this system is ${\bf w}=0$. Therefore, the macroscopic operator for the generalized momentum transport equation, obtained through MFM, is a correct RANS operator in the sense that it admits the true RANS solution. 

\begin{figure}
%\vspace{4 mm}
\includegraphics[width=0.98\textwidth]{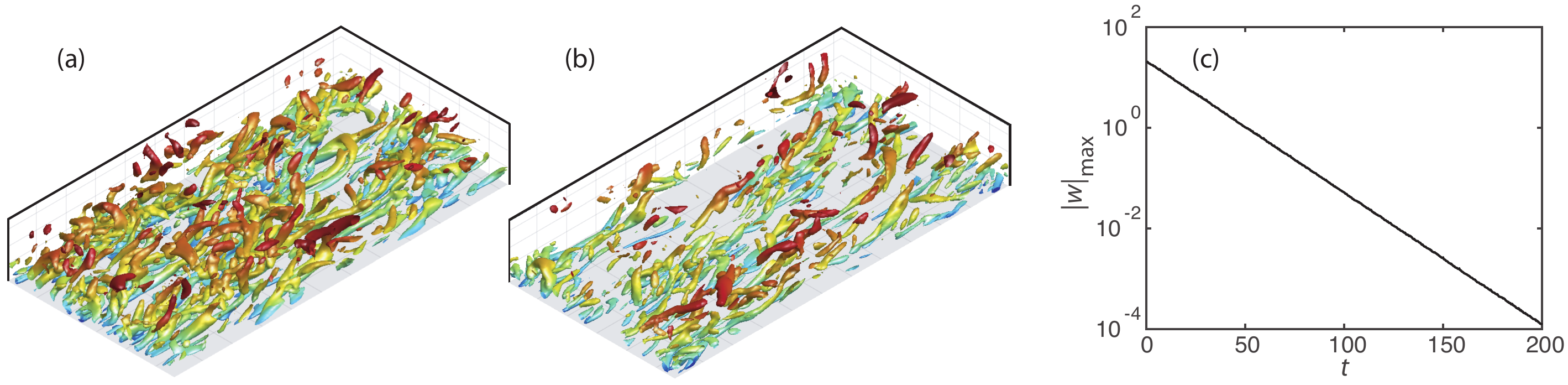}
\caption{Instantaneous iso-contour of the Q-criterion at $t=0$ (a) and $t=5\delta/u_\tau$ (b) for ${\bf w}={\bf v}-{\bf u}$ representing the difference between the NS solution to a turbulent channel flow at Re$_\tau=180$ and a GMT solution with matching boundary condition and forcing.  The initial condition for GMT is ${\bf v}={\bf 0}$. Maximum  magnitude of the difference field, $|{\bf w}|_\text{max}$ versus time is shown in (c).}
\label{fig:fig8}
\end{figure}

The decay rate of field ${\bf w}$, indicates an important time scale that needs to be considered in MFM. When performing averaging of the microscopic solutions, one should consider time windows much larger than this decay time as a condition for convergence of statistics and independence to the initial conditions. In practice, ${\bf w}$ decays exponentially after an early transient. As an example, we performed simulations of a turbulent channel flow at $\text{Re}_\tau=180$, and examined the solution to the GMT, Equation~(\ref{eq:gns}), when it is supplemented with the same macroscopic forcing and boundary conditions as those of the NS solution but using an arbitrary initial condition. Figure~\ref{fig:fig8} shows the computed norm of ${\bf w}$ versus time confirming its exponential decay. A fitted line to this plot suggests a decay time scale of about $\tau_\text{mix}=16.6 \delta/u_\tau$, where $\delta$ is the channel half width, and $u_\tau$ is the shear scale velocity. Practically, $\tau_\text{mix}$ is the time scale that denotes slowest mode of mixing and can be used as a mixing index; systems with smaller $\tau_\text{mix}$ are better mixers. Generally, $\tau_\text{mix}$ for mixing of scalars can be different than those for mixing of momentum as suggested by the results shown in~\cite{Shirian2019}. 

One novel aspect of this work is probably recasting the Navier-Stokes equation in terms of a generalized momentum transport equation, which we called GMT~(\ref{eq:gns}). This generalization formalizes two interpretations of velocity as a vector quantity. The first interpretation is a kinematic interpretation; it represents the time rate of material volume crossing an interface per unit area. The second interpretation of velocity is a dynamic one; it represents momentum per unit mass which is the ability of the material to resist against forces. These two interpretations are not new. They are in fact emphasized often in graduate teaching of momentum analysis in terms of the Reynolds Transport Theorem. In RANS modeling, the goal is often prediction of forces, and thus one thinks of the dynamic interpretation of velocity. The mission of RANS models is therefore determining how the mean momentum is being transported by the underlying turbulent flow. When dealing with the nonlinear advective term, $\partial/\partial x_j\left(u_j u_i\right)$, the first velocity, $u_j$ is the kinematic velocity; it does not represent momentum, but it is a geometric ``transporter" of momentum. The second velocity, $u_i$, is the quantity of interest; it is the ``transportee," representing momentum per unit mass. The RANS problem can be recast into the following question: how does turbulence mix momentum? 

Answer to this question is dependent on the underlying turbulence. We therefore need to first specify the turbulent flow field of interest. This is done here by solving the original NS as the ``donor" equation, whose solution ``donates" a quantitative representation of the turbulence field of interest. The GMT equation is the ``receiver" system that focuses specifically on the donated turbulence field and determines quantitatively how that field transports momentum in general. MFM analysis of GMT then provides an assessment of mean momentum transport when projected to the mean space.

In systems governed by the Navier-Stokes equation, the transporter and transportee fields are further constrained to be equal. However, this is only a special case of GMT. By obtaining an operator representing GMT, we also cover the special case of NS.  

While MFM uses linear techniques, the final macroscopic RANS equation is indeed a nonlinear equation; it involves an operator acting linearly on the velocity but the operator itself is dependent on the donated velocity. The caveat is that MFM obtains the macroscopic operator in terms of the independent coordinates for a given flow. An analytical MFM solution would allow expression of the macroscopic operator directly in terms of the statistics of the underlying flow and would resolve the RANS closure problem. Numerical MFM solutions, however, fall short in accomplishing this final goal. Nevertheless, numerical MFM allows substantial advancement towards achieving this goal by determining the RANS operators in terms of independent coordinates and thus quantifying model-form errors in existing RANS models. 
%In other words, numerical MFM acts as an advanced rheometer determining how a given flow mixes momentum. Whether this mixing is describable by local mean gradients or non-local operators, and whether the tensorial coefficients are isotropic or there are substantial non-Boussinesq effects, MFM reveals these details quantitatively. 
One hope is that much of these missing pieces in current RANS models be universal and thus application of MFM on canonical problems could inform models that will be predictive of practical scenarios.

Due to the limited scope of this report, we skip providing examples of the application of MFM for analysis of turbulent flows. However, this is our main intention in developing the presented methodology. %In a separate work~\cite{Shirian2019} we have applied the same methodology presented here to study more complex flows involving 3D turbulent unsteady velocity fields. In this context we find that the large-eddy size in turbulence plays a role analogous to the mean free path, and thus controls the mixing length. For macroscopic scales on the order of or smaller than the large eddy, macroscopic operators depict considerable non-local effects, which are missing in the mainstream eddy diffusivity closure models. An additional issue is anisotropy of the macroscopic operators in multi-dimensions which persists even when the macroscopic scale is large.  %As an example, we show in~\cite{Shirian2019} how application of MFM in homogeneous and isotropic turbulence can inform new RANS operators which then led to significant improvements in prediction of axisymmetric jet flows without requiring an MFM analysis of jets. 

\section{Discussions}
We presented a systematic numerical procedure, which we call MFM, for determination of scaled up operators describing the macroscopic influence of microscopic transport processes. MFM is applicable to both laminar and turbulent flows, homogeneous and inhomogeneous flows, and can be used to examine transport of scalar fields as well as vector fields such as those describing momentum as a transported quantity. 

For a given setting, MFM can obtain the exact macroscopic differential operator governing the evolution of mean fields. %However, MFM does not solve the closure problem in turbulence modeling. This is because MFM obtains the operator coefficients explicitly in terms of the independent coordinates for each specific setup. Tackling the turbulence closure question requires expressing these coefficients in terms of the mean field quantities themselves resulting in nonlinear operators.  The advantage of MFM is therefore in quantification of the model-form error and identification of improved model forms. Whether the RANS model should involve an anisotropic but local eddy diffusivity or a non-local operator, the presented methodology can provide this information quantitatively. 
Additionally, MFM provides a robust framework for RANS model verification. In today's practice, RANS model forms are postulated based on physical intuition, and their coefficients are tuned using reference data from configurations similar to those in real applications. Therefore, when a model generates accurate mean velocity profiles, it is unclear to what degree the good performance is due to cancellation of model-form error with coefficient error. Performing MFM on specific geometries, and direct comparison with RANS operators allows a more rigorous verification of RANS models.  

%We point out that MFM is an expensive procedure (either forward or inverse), involving at least multiple DNS solutions. Therefore, we expect this method to be mostly used to investigate canonical problems. The hope is that an understanding of closure operators in canonical flows will reveal some universality, and that this understanding can be extendable to flows with more complex geometries. 

Examination of the macroscopic operators for homogeneous flows in Fourier space revealed a smooth curve with $k^2$ scaling in the low-wavenumber limit and $k^1$ scaling in the high-wavenumber limit. The $k^1$ scaling was universally observed in a wide range of conditions for both laminar and turbulent problems and for both scalar and momentum transport. While the $k^1$ scaling is consistent with the intuitive expectation of bounded dispersion speed, the pre-factor to this scaling is real and positive, and thus cannot be captured by an advective operator. Instead, we showed that this scaling is indicative of a non-local, but dissipative mechanism of transport, as suggested by the fitted operator, $\nabla \left[D/\sqrt{1-l^2\nabla^2}\right]\nabla$. Due to presence of two asymptotic scalings in the small and large $k$ limits, the non-locality does not conform to a simple fractional-order differential operator expressed as a Laplacian to a non-integer power. Extending our observations to inhomogeneous cases, we conclude that fractional order PDEs are likely limited in scope for describing macroscopic transport. Instead one needs to consider non-locality in a broader sense, as described by the introduced kernel $\mathcal{D}\left({\bf x},{\bf y}\right)$ that expresses mean fluxes in terms of a weighted superposition of the surrounding mean field gradients. 

 A useful complementary study to this work would be the development of fast and low-memory computation of non-local integral operations. Ideally, to save time in both MFM and RANS calculations one should aim to utilize fast convolutions whose moments can be systematically informed by IMFM. 

 We acknowledge that the introduced eddy diffusivity operators in generic form as kernel integrals in Equation~(\ref{eq:1dkernel}) has been recognized in previous studies in broad sense. In this context, specifically the work of Hamba\cite{Hamba2004,Hamba2005} is closest to the work presented here. However, our work involves major differences in methods of obtaining the closure operators as well as in the results developed. Specifically, the present work offers a systematic computational framework for determination of closure operators with an additional flexibility for substantial cost reduction based on method of measuring kernel moments using IMFM. Additionally, our results provided eddy diffusivity not only in terms of non-local integrals, but also we utilized the Fourier representation to arrive at expressing these operators in compact stand-alone analytical PDE form. The diffusivity operator $\mathcal{D} = D/\sqrt{1-l^2\nabla^2}$, presented for homogeneous systems, is an example of compact PDE representation of a non-local macroscopic closure which we derived and quantitatively validated. Furthermore, the present work provides a harmonious connection between closure modeling of scalar transport and that for momentum transport by extending MFM in the context of the Generalized Moment Transport equation, with provable convergence to the Navier-Stokes solution. %Lastly, the pedagogical examples involving parallel flows and connection to Taylor dispersion, presented here, provide additional insights into macroscopic closure and broader scope for utilization of the tools developed.

 It might also be possible to apply MFM to the problem of LES modeling in the long term. One fundamental approach is to recast the LES modeling problem in the MFM context by modifying the definition of averages that are used here. %A more immediately accessible approach is to utilize knowledge derived from MFM in the RANS context. For example, the reported $k^1$ scaling of macroscopic operator for the high-wavenumber limit is also expected in the case of LES, implying the LES eddy diffusivity should vanish in the high-wavenumber limit. %Additionally, the required anisotropy treatments in wall-modeled LES can be informed from anisotropy of $\mathcal{D}$ to be understood in the RANS context. 

% While in this report we considered scalar and momentum transport in the context of incompressible flows, it maybe possible to envision extension of MFM for broader applications including reacting scalars, compressible systems, shock-turbulence interactions, and buoyancy driven flows. These are examples where anisotropy and/or coupling with other physics is known to impose challenge in modeling. MFM is a valuable tool that can shed light into effects of multiphysics coupling on closure operators in a quantitative manner. 

Lastly, MFM can be viewed as a numerical rheometer, similar to the way that molecular dynamics simulations predict continuum-level transport coefficients. Experimental rheometers measure momentum diffusivity due to the underlying chaotic Browning dynamics. When using them, the key fundamental assumption is that placing of the material under the rheometer does not affect the chaotic Browning dynamics responsible for transport of momentum. Experimental rheometry of turbulence is not possible due to violation of this condition. In this sense, a novel aspect of our work is recasting the Navier-Stokes equation in a more general form by separating the roles of velocity as a transporter and transportee. Rigorous rheometry must be non-intrusive to the transporter mechanism but it can be intrusive to the transportee field. Given emerging supercomputing power, MFM provides a fresh opportunity to study turbulence by measuring its ``material properties". 

\section{Acknowledgements}
This work was supported by the Boeing Company under grant number SPO 134136 and by the National Science Foundation under award number 1553275. D.P. was also supported by a Kwanjeong Educational Foundation Fellowship. The authors are grateful for critical comments from Drs. Chad Winkler, Philippe Spalart, and Mori Mani through the course of the research resulting in this report. The authors are in debt to Jessie Liu and Dr. Karim Shariff for their technical discussions and elaborate comments on a draft of this report. The phrase ``Generalized Momentum Transport" (GMT) as a name for Equation~(\ref{eq:gns}) was suggested in discussion with Dr. Karim Shariff. The terminologies of ``donor" and ``receiver", in explaining MFM for momentum transport were suggested by Dr. Brandon Morgan. We are grateful to Prof. Tamer Zaki for pointing the connection between this work and the work of Dr. Fujihiro Hamba. A.M. is grateful to Dr. Parvin Mani for hosting him in San Diego during his sabbatical leave where a foundational part of this work was developed. 

\appendix
\section{Solution to Equation~(\ref{eq:c_0})}
\label{ap:solution}
Substitution of Equation~(\ref{eq:series}) into~(\ref{eq:c_0}) results in a series involving $\cos\left(nx_2\right)$. The product  $\cos\left(nx_2\right)\cos(x_2)$ can be expanded as $(1/2)\cos\left[\left(n+1\right)x_2\right]+(1/2)\cos\left[\left(n-1\right)x_2\right]$. Using this expansion and balancing the left-hand side and right-hand side of the equation for each term results in
\begin{align}
i\omega a_0+\frac{ik}{2}a_1=1, \label{eq:n0}\\
ika_0+(i\omega+1)a_1+\frac{ik}{2}a_2=0, \label{eq:n1}\\
\frac{ik}{2}a_{n-1}+(i\omega+n^2)a_n+\frac{ik}{2}a_{n+1}=0,
\label{eq:recursive}
\end{align}
where the last equation applies for $n\ge2$. The recursive relation in~(\ref{eq:recursive}) needs two initial conditions, i.e., $a_1$ and $a_2$. Depending on the ratio of $a_1$ and $a_2$, it admits two modes of solutions, one increasing in magnitude exponentially, and one decreasing in magnitude exponentially. To obtain a physical solution, we impose the condition that the exponentially increasing mode must be zero. This condition sets a unique value for the ratio $a_1/a_2$. We investigated this condition by rewriting~(\ref{eq:recursive}) in terms of $r_n=a_{n}/a_{n-1}$ and re-arranging as
$$
 r_{n+1}=\frac{2\left(-n^2-i\omega\right)}{ik}-\frac{1}{r_n}.
$$
The question is to determine the value of $r_2$ such that this recursive relation leads to $|r_{n\rightarrow\infty}|$ smaller than unity. A randomly selected $r_2$ leads to unbounded $|r_{n\rightarrow\infty}|$. This can be understood better when thinking in terms of Equation~(\ref{eq:recursive}). A random combination of $a_1$ and $a_2$ will have both exponentially increasing and decreasing modes present. Even with a small pre-factor, the exponentially increasing mode eventually prevails for sufficiently large $n$. Only when the ratio of $a_1$ and $a_2$ are set to precisely remove the increasing mode, one will obtain the physically correct solution. Our numerical trick is to use the recursive relation backward in $n$. By choosing a random finite number for $r_{N}$ where $N$ is a sufficiently large number, we solve the above recursive relation backward in $n$ to find a value for $r_2$. Moving backward in $n$ suppresses the relative contribution of the exponentially increasing mode. Starting from sufficiently large $N$, on the order of $~O(100)$, we find the physical value of $r_2$ to be converged within many significant digits. Once the ratio $a_2/a_1$ is at hand, one can solve Equations~(\ref{eq:n0}) and~(\ref{eq:n1}) to obtain $a_0$. 

\section{Numerical Solution of Equation~(\ref{eq:approxL})}
\label{ap:numerical}
Even though in other sections we have solved macroscopic equations directly in physical space using techniques of linear algebra, for the specific problem discussed in Section~\ref{sec:taylorrevisit}, it was easier to solve the problem in the Fourier space for both spatial and temporal dimensions. To do this, we represent the initial condition as a source term that is activated at time $t=0$ via a Dirac delta function, 
$$
s(x_1,t)=\exp\left(-x_1^2/0.025\right)\delta(t).
$$
The above expression is Fourier transformed in space and time using a physical mesh with $\Delta x_1=0.05$ and $\Delta t=0.025$ and a truncated domain $-5\le x_1 \le 5$. $\widehat{s}\left(k,\omega\right)$ is then divided by $\widehat{\overline{\mathcal{L}}}\left(k,\omega\right)=\sqrt{\left(1+i\omega\right)^2+k^2} - 1$ and then inverse transformed to obtain $\overline{c}\left(x_1,t\right)$ in the physical space. 

Next, we discuss the treatment of mode $k=\omega=0$. Given that the temporal domain is not infinite, this procedure results in a time-periodic solution corresponding to repeated spontaneous injections of scalar in the domain. To eliminate the artifacts associated with this periodicity, we applied the following remedies. First, the temporal domain is taken to be substantially large ($0\le t \le 40$). This time is longer than the diffusion time over the entire spatial domain, and ensures that for each period the effects of previous cycles have vanished except for their spatial mean. The second remedy is to correct the spatial mean of $\overline{c}$ to avoid a linear growth due to the periodic cycles. To this end, the mode $k=0$ is solved analytically in the physical space using the trivial differential equation $\partial/\partial t =0$. 

To verify our method, we repeat this procedure using the exact $\widehat{\overline{\mathcal{L}}}$, obtained through MFM discussed in Appendix~\ref{ap:solution} and confirm that the obtained $\overline{c}(x_1,t)$ matches that of the two-dimensional DNS solved in the physical space under resolution refinement.

\section{Comparison against solution to the telegraph equation}
\label{ap:telegraph}
\begin{figure}[h]
%\vspace{4 mm}
\includegraphics[width=0.42\textwidth]{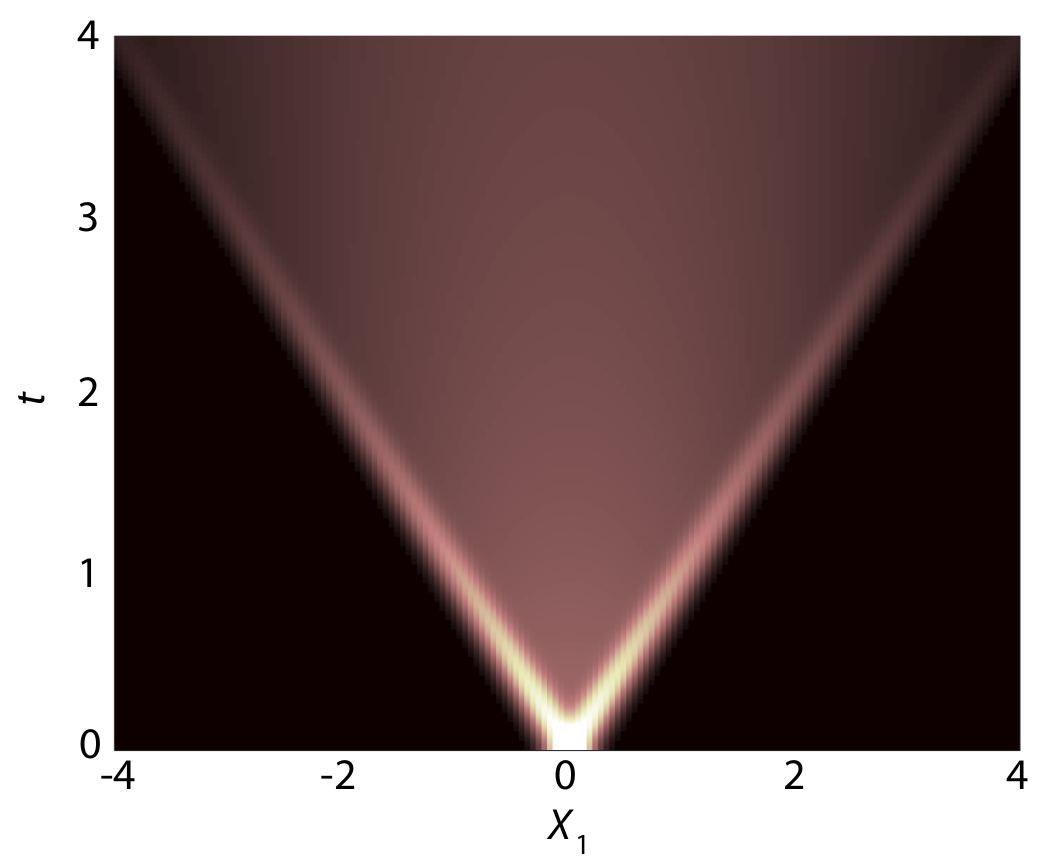}
\caption{Space-time evolution of $x_2$-averaged concentration field with an initial condition $c=\exp\left(-x_1^2/0.025\right)$ subject to the telegraph equation}
\label{fig:fig9}
\end{figure}

Following a request by anonymous referee 1, we present here the prediction of the telegraph equation for the solution to the same problem introduced in figures~(\ref{fig:fig1}) and~(\ref{fig:fig2}). Figure~(\ref{fig:fig9}) shows this solution which maybe directly compared to panels (a) through (d) in Figure~(\ref{fig:fig2}). Although this result shows finite dispersion speed at early times, indicating a performance better than the Taylor's solution, it lacks in terms of providing key features of the solution, namely around the “edge” of the concentration profile. 

The appropriate form of the telegraph equation in this case is
$$
\frac{\partial c}{\partial t}+\frac{1}{2}\frac{\partial^2c}{\partial t^2}=\frac{1}{2}\frac{\partial^2c}{\partial x^2} + s(x_1,t).
$$
Coincidentally, in the absence of the source term, $s$, one can show that our model presented by Equation~(\ref{eq:approxLphys}) can be simplified to the above equation. However, as shown in the computational results, our model provides a significantly better solution in comparison in to the telegraph equation. The reason is that the source term cannot be ignored when manipulating models. Specifically in this case, the initial condition essentially appears as a source term with Dirac delta distribution in time and with a non-local influence on the spatio-temporal evolution of the solution.

\section{Kernel Reconstruction from Moments}
\label{ap:kernel}
For the problem considered in Section~\ref{sec:construct}, we considered an approximate kernel described by 
\begin{equation}
\mathcal{D}(x_1,y_1) = \left\{
  \begin{array}{lr}
    h\left(x_1\right)\exp\left(\frac{x_1-y_1}{l_1\left(x_1\right)}\right)& , x_1 < y_1 ;\\
    h\left(x_1\right)\exp\left(\frac{y_1-x_1}{l_2\left(x_1\right)}\right) & , x_1 \ge y_1 .
  \end{array}
\right.
\end{equation}
For each $x_1$-position, the moments of the above kernel form can be computed analytically in terms of $h$, $l_1$ and $l_2$. By matching these moments to those obtained from IMFM, we obtained profiles of $h$, $l_1$ and $l_2$ versus $x_1$. Aside from kernel shape approximation and discretization errors, we commit two additional errors when matching the moments. First, the analytical computation of the moments in terms of $h$, $l_1$ and $l_2$ does not consider domain truncation by walls. In other words, we assume that $l_1$, $l_2$, and $h$ become sufficiently small near the wall to allow neglecting of this error. Second, in rare points where the matched value of $l_1$ or $l_2$ becomes negative, we replace its value with zero. By utilizing such kernels we show in Figure~\ref{fig:fig7} significant improvements in prediction of mean concentration field in comparison to that obtained from the local eddy diffusivity. This kernel reconstruction method also resolves the convergence issue observed when corrections to the leading operator was incorporated using expansions based on Taylor series and moments.  

For the specific problem considered here, we avoided using the canonical homogeneous kernel, $\mathcal{D}=D/\sqrt{\mathcal{I}-l^2\partial^2/\partial x_1^2}$ because this kernel is symmetric in space, while the actual kernel involves spatial asymmetry near the boundaries. However, we here briefly note that in a homogeneous system the moments of this kernel are $D^0=D$, $D^1=0$, and $D^2=Dl^2/2$. 
\bibliographystyle{unsrt}
\bibliography{paper_vs3}

\end{document}